\documentclass[preprint]{aastex63}

\shorttitle{Cosmic chronometers and spatial curvature}
\shortauthors{Vagnozzi \textit{et al.}}

\begin{document}

\title{Eppur \`{e} piatto?\\The cosmic chronometer take on spatial curvature and cosmic concordance}

\author[0000-0002-7614-6677]{Sunny Vagnozzi}
\affiliation{Kavli Institute for Cosmology, University of Cambridge, Madingley Road, Cambridge CB3 0HA, United Kingdom}
\correspondingauthor{Sunny Vagnozzi}
\email{sunny.vagnozzi@ast.cam.ac.uk}

\author[0000-0003-4330-287X]{Abraham Loeb}
\affiliation{Department of Astronomy, Harvard University, 60 Garden Street, Cambridge, MA 02138, USA}

\author[0000-0002-7616-7136]{Michele Moresco}
\affiliation{Dipartimento di Fisica e Astronomia ``Augusto Righi'',\\Alma Mater Studiorum Universit\`{a} di Bologna, via Piero Gobetti 93/2, I-40129 Bologna, Italy}
\affiliation{INAF - Osservatorio di Astrofisica e Scienza dello Spazio di Bologna, via Piero Gobetti 93/3, I-40129 Bologna, Italy}

\newcounter{appendix}
\renewcommand{\theappendix}{A}

\begin{abstract}

\noindent The question of whether Cosmic Microwave Background (CMB) temperature and polarization data from \textit{Planck} favor a spatially closed Universe with curvature parameter $\Omega_K<0$ has been the subject of recent intense discussions. Attempts to break the geometrical degeneracy combining \textit{Planck} data with external datasets such as Baryon Acoustic Oscillation (BAO) measurements all point towards a spatially flat Universe, at the cost of significant tensions with \textit{Planck}, which make the resulting dataset combination problematic. Settling this issue requires identifying a dataset which can break the geometrical degeneracy while not incurring in these tensions. We argue that cosmic chronometers (CC), measurements of the expansion rate $H(z)$ from the relative ages of massive early-type passively evolving galaxies, are the dataset we are after. Furthermore, CC come with the additional advantage of being virtually free of cosmological model assumptions. Combining \textit{Planck} 2018 CMB temperature and polarization data with the latest CC measurements, we break the geometrical degeneracy and find $\Omega_K=-0.0054 \pm 0.0055$, consistent with a spatially flat Universe and competitive with the \textit{Planck}+BAO constraint. Our results are stable against minimal parameter space extensions and CC systematics, and we find no substantial tension between \textit{Planck} and CC data within a non-flat Universe, making the resulting combination reliable. Our results allow us to assert with confidence that the Universe is spatially flat to the ${\cal O}(10^{-2})$ level, a finding which might possibly settle the ongoing spatial curvature debate, and lends even more support to the already very successful inflationary paradigm.

\end{abstract}

\keywords{cosmic background radiation --- cosmological parameters --- cosmology: observations --- distance scale --- galaxies: general}

\section{Introduction}
\label{sec:intro}

The question concerning what is the shape of the Universe, or more precisely the local geometry of the observable Universe, is one of central importance in cosmology. This question can be addressed by determining the spatial curvature of the Universe (hereafter simply curvature), a quantity which specifies how much the Universe's local geometry differs from flat space geometry. In practice, this usually amounts to measuring or constraining the curvature parameter $\Omega_K$, with $\Omega_K=0$ corresponding to a spatially flat Universe, whereas $\Omega_K<0$ and $\Omega_K>0$ correspond to a spatially closed and spatially open Universe respectively. The curvature parameter quantifies the effective contribution of spatial curvature to the energy density of the Universe today.~\footnote{In principle $\Omega_K$ could vary as a function of time, if resulting from large-scale density inhomogeneities produced during an early superhorizon process such as inflation or alternative scenarios. In this case, the evolving value of $\Omega_K$ reflects the fact that our causally connected region samples an evolving volume. However, the value of an evolving $\Omega_K$ should be very small, else we should already have detected its signature, \textit{e.g.} in the late-integrated Sachs-Wolfe effect.}

The importance of obtaining high-fidelity constraints on $\Omega_K$ cannot be overstated. The sign and value of $\Omega_K$ play an important role in determining the future evolution of the Universe. From the model-building side, constraints on $\Omega_K$ have important consequences for models of inflation, most of which predict an Universe which is spatially flat to the level of $\vert \Omega_K \vert \lesssim {\cal O}(10^{-4})$~\citep[see e.g.][]{Kazanas:1980tx,Starobinsky:1980te,Guth:1980zm,Sato:1981ds,Mukhanov:1981xt,Linde:1981mu,Albrecht:1982wi}. Conversely, detecting $\vert \Omega_K \vert \neq 0$ at the ${\cal O}(10^{-2})$ level or larger could be a problem for most models of inflation~\citep{Linde:2007fr,Kleban:2012ph,Guth:2012ww}, although others~\citep[see e.g.][]{Bull:2013fga} have argued that this might not be as problematic. It is generally simpler to construct inflationary models in a spatially open Universe~\citep[see e.g.][]{Coleman:1980aw,Gott:1982zf,Ratra:1994ni,Ratra:1994vw,Ratra:1994dm,Bucher:1994gb,Linde:1995xm,Yamamoto:1995sw,Linde:2007fr,Kleban:2012ph,Guth:2012ww}, whereas achieving the same result in a spatially closed Universe might require more fine-tuning~\citep[see e.g.][]{Ratra:1984yq,Hartle:1983ai,Linde:2003hc,Ratra:2017ezv}. In any case, the importance of spatial curvature in modern cosmology is the reason why a huge number of works have been devoted to providing and forecasting constraints on $\Omega_K$ from current and future cosmological observations.~\footnote{For an inevitably incomplete list of such works, see e.g.~\cite{Vardanyan:2009ft,Carbone:2011bx,Li:2012vn,Bull:2014rha,Takada:2015mma,DiDio:2016ykq,Leonard:2016evk,Rana:2016gha,Ooba:2017npx,Jimenez:2017tkk,Ooba:2017lng,Park:2017xbl,Denissenya:2018zcv,Park:2018bwy,Park:2018fxx,Park:2018tgj,Bernal:2018myq,Li:2019qic,Park:2019emi,Wang:2019yob,Zhai:2019nad,Geng:2020upf,Heinesen:2020sre,Gao:2020irn,Khadka:2020hvb,Nunes:2020uex,Liu:2020pfa,Chudaykin:2020ghx,Benisty:2020otr,Shimon:2020dvb,Troster:2020kai,DiValentino:2020kpf,Qi:2020rmm}.}

Up to the early 2010s, it had been the case that cosmological observations were unquestionably consistent with the Universe being spatially flat to within the then current precision~\citep[e.g.][]{Bennett:1996ce,Melchiorri:1999br,deBernardis:2000sbo,Balbi:2000tg,Hinshaw:2012aka}. On the other hand, whether this conclusion still holds in light of the latest measurements of Cosmic Microwave Background (CMB) temperature and polarization anisotropies from the \textit{Planck} satellite 2018 legacy data release~\citep[][P18 hereafter]{Akrami:2018vks,Aghanim:2018eyx,Aghanim:2019ame} is not completely clear. At face value, these measurements \textit{appear} to favor a spatially closed Universe at the ${\cal O}(10^{-2}-10^{-1})$ level, with $-0.095<\Omega_K<-0.007$ at 99\%~confidence level (C.L.), a figure which one could be tempted to interpret as a genuine detection of $\Omega_K \neq 0$, and which could spell huge trouble for the otherwise extremely successful inflationary paradigm.

Caution is required before jumping to the previous conclusion, for at least two important reasons. The first reason is that the P18 preference for $\Omega_K<0$ is driven to an important extent by the anomalous preference for extra gravitational lensing-induced smoothing in P18's temperature higher order acoustic peaks. In addition, $\Omega_K<0$ allows for a better fit to a number of anomalously low features in the low-$\ell$ CMB multipoles~\citep{Efstathiou:2003hk}. The lensing anomaly is captured through the phenomenological parameter $A_L$~\citep{Calabrese:2008rt}, which artificially rescales the CMB power spectra lensing amplitude. The preference for extra lensing is reflected in $A_L>1$~\citep{Aghanim:2018oex}. Is this preference real or simply a fluke? If the latter interpretation is the correct one, the preference for $A_L>1$ (and correspondingly $\Omega_K<0$) should decrease when more data (or, in the case of the CMB, a larger sky fraction) is used. The recent reanalysis of the \textit{Planck} High Frequency maps by~\cite{Efstathiou:2019mdh}, with access to a larger sky fraction (but discarding the $100 \times 100\,{\rm GHz}$ spectrum) thanks to a modified version of the \texttt{CamSpec} likelihood, does indeed go in this direction. Nonetheless, a preference for $\Omega_k \neq 0$ and $A_L \neq 1$, albeit at a lower statistical significance, remains~\citep{Efstathiou:2019mdh}. It is also worth mentioning that the latest Atacama Cosmology Telescope (ACT) DR4 results do not show any indication for a lensing anomaly, and are consistent with $\Omega_K=0$ and $A_L=1$~\citep[see][]{Aiola:2020azj}.

The second reason why caution is required is that, when it comes to spatial curvature, the constraining power and reliability of CMB temperature and polarization anisotropies is limited by the so-called geometrical degeneracy~\citep{Bond:1997wr,Zaldarriaga:1997ch,Efstathiou:1998xx}. We will return to this important issue later on, but in essence the geometrical degeneracy reflects the fact that certain key cosmological parameters can be re-arranged in combinations which keep the CMB temperature anisotropy power spectrum largely unchanged. The geometrical degeneracy notably affects the matter density parameter $\Omega_m$, the Hubble constant $H_0$, and the curvature parameter $\Omega_K$. In order to stabilize constraints on $\Omega_K$ coming from CMB temperature and polarization anisotropies data alone, it is vital to combine these with ``external'' measurements which are able to break the geometrical degeneracy. This is essentially achieved by constraining the late-time expansion history and/or pinning down at least one parameter between $\Omega_m$, $H_0$, and $\Omega_K$.

A limited set of examples of external measurements one can use to break the geometrical degeneracy includes but is not limited to the CMB lensing power spectrum reconstructed from the temperature 4-point function (in this case the ``external'' qualifier is used rather loosely), Baryon Acoustic Oscillation (BAO) distance and expansion rate measurements, full-shape (FS) galaxy power spectrum measurements, Type Ia Supernovae (SNeIa) distance moduli measurements, and local measurements of the Hubble constant $H_0$, for instance from Cepheid- or Tip of the Red Giant Branch-calibrated SNeIa or from strongly lensed quasars. Combining P18 data with these datasets, either one or more at a time, within a minimal 7-parameter $\Lambda$CDM+$\Omega_K$ model, breaks the geometrical degeneracy and delivers constraints consistent with $\Omega_K=0$~\citep{Aghanim:2018eyx,Efstathiou:2020wem,Vagnozzi:2020zrh}.~\footnote{When SNeIa distance moduli or local $H_0$ measurements are considered, the assumption of a minimal 7-parameter $\Lambda$CDM+$\Omega_K$ model is crucial, as freeing up the dark energy equation of state $w$ can significantly alter the conclusions and push the constraints towards $\Omega_K<0$ once more, as shown in~\cite{DiValentino:2020hov}.} The issue here is that, \textit{within the assumption of a curved Universe}, P18 is in tension with each and every one of these external probes. Constraints arising from datasets in tension within a given model should at the very least be viewed with suspicion. In other words, P18 data cannot confidently be combined with these external datasets (even though the latter are crucial to break the geometrical degeneracy) as long as this tension persists. The latter view was strongly upheld by~\cite{Handley:2019tkm} and~\cite{DiValentino:2019qzk}, where the aforementioned tensions were rigorously quantified, with their significance found to be between the $2.5\sigma$ and $4.5\sigma$ level, depending on the specific dataset combination as well as tension metric considered.

Besides these two important concerns, there is a third somewhat minor issue affecting at the very least BAO and FS measurements. As we will discuss in more detail later, the data reduction process leading to these measurements requires at more than one point making a fiducial cosmology assumption, with the choice falling on $\Lambda$CDM. The question is then whether this residual model-dependence poses a problem in the interpretation of the results obtained combining these datasets with P18 data, particularly given the rather extreme values of $\Omega_K$ suggested by P18 alone. These questions were partially addressed in e.g.~\cite{Ding:2017gad,Sherwin:2018wbu,Carter:2019ulk,Heinesen:2019phg,Bernal:2020vbb}, although it is fair to say that a definitive conclusion on the matter has yet to be reached.

With these three important issues in mind, it would appear that we have reached an impasse in the spatial curvature conundrum. We clearly need to stabilize P18's constraints on $\Omega_K$ by combining P18 with an external dataset (ideally carrying the least amount of model-dependent assumptions as possible) to break the geometrical degeneracy, but this combination should not be in tension within a $\Lambda$CDM+$\Omega_K$ cosmology. One way to exit this impasse and shed more light on the spatial curvature conundrum is therefore to find a ``golden dataset'' which helps to break the geometrical degeneracy once combined with P18, is not in strong tension with P18 when assuming a curved Universe, and ideally carries the least possible amount of model-dependent assumptions. 

Does such a ``golden dataset'' exist and if so what does it tell us about the spatial curvature of the Universe? In this paper, we will find that the answer to the first part of the question is ``Yes''. In fact, we shall argue that cosmic chronometers (sometimes also referred to as cosmic clocks), \textit{i.e.} measurements of the expansion rate $H(z)$ from the relative ages of passively evolving galaxies, satisfy all three the requirements outlined above, while also having two additional minor advantages over the external datasets we mentioned previously. The principle underlying the use of cosmic chronometer data to measure $H(z)$ was proposed by one of us nearly 20 years ago, in~\cite{Jimenez:2001gg}. Combining P18 with cosmic chronometer data, we will find that the answer to the second part of the previous question is instead that the Universe appears to be spatially flat given the achievable sensitivity to $\Omega_K$. The latter is worse by only a factor of $\approx 2.5$ compared to the sensitivity to $\Omega_K$ obtained from the P18+BAO dataset combination.

The rest of this paper is then organized as follows. Section~\ref{sec:curvatureandchronometers} is an introductory section, with Section~\ref{subsec:curvaturegeometrical} focused on reviewing the role and importance of spatial curvature in modern cosmology, as well as discussing in more depth the geometrical degeneracy issue, whereas Section~\ref{subsec:cc} reviews the principle underlying the cosmic chronometer measurements. In Section~\ref{sec:methods} we discuss the statistical methods and observational datasets we make use of. Our results are discussed in Section~\ref{sec:results}, with Section~\ref{subsec:tension} devoted to quantifying the tension between \textit{Planck} and cosmic chro nometer data within a curved Universe, Section~\ref{subsec:ageofoldest} discussing our results in light of the ages of the oldest objects in the Universe, and Section~\ref{subsec:extended} assessing the stability of the previous results against extended parameter spaces. Finally, in Section~\ref{sec:conclusions} we draw concluding remarks. Appendix~\ref{sec:appendix} assesses the impact on our results of observational systematics affecting the cosmic chronometer measurements.

\section{Spatial curvature, the geometrical degeneracy, and cosmic chronometers}
\label{sec:curvatureandchronometers}

In this Section we clarify our notation and briefly review the role of spatial curvature in modern observational cosmology. We then revisit the geometrical degeneracy present in CMB data, which affects CMB-only constraints on the curvature parameter $\Omega_K$. Finally, we discuss the use of cosmic chronometers to map the late-time expansion history of the Universe, and how these measurements can help break the geometrical degeneracy once combined with CMB data.

\subsection{Spatial curvature and the geometrical degeneracy}
\label{subsec:curvaturegeometrical}

Our backbone assumptions in describing the Universe on large scales are General Relativity (GR) and the cosmological principle. In reduced spherical polar coordinates $(r,\theta,\phi)$ and denoting cosmic time by $t$, on sufficiently large scales the Universe is described by the Friedmann-Lema\^{i}tre-Robertson-Walker (FLRW) metric, with line element given by:
\begin{eqnarray}
ds^2 = -dt^2+a^2(t) \left [ \frac{dr^2}{1-Kr^2} + r^2 \left ( d\theta^2+\sin^2\theta d\phi^2 \right ) \right ]\,,
\label{eq:flrw}
\end{eqnarray}
where the scale factor $a(t)$ describes the expansion/contraction of homogeneous and isotropic spatial slices as a function of time. The spatial slices have constant spatial curvature determined by the parameter $K$. Negative spatial curvature $K=-1$ corresponds to open hyperbolic space, positive spatial curvature $K=+1$ to closed hyperspherical space, and vanishing spatial curvature $K=0$ to flat Euclidean space. At the level of the Friedmann equations, spatial curvature effectively contributes an additional matter-energy source, with fractional contribution quantified by the curvature parameter $\Omega_K \equiv -K/(H_0a_0)^2$. Here the Hubble parameter $H$ characterizes the expansion rate of the Universe, and $_0$ denotes quantities evaluated today: $H_0$, the Hubble parameter evaluated today, is typically referred to as the Hubble constant. Importantly, $\Omega_K$ and $K$ come with opposite signs, so that an open Universe corresponds to $\Omega_K>0$ and a closed Universe to $\Omega_K<0$.

Determining $\Omega_K$ from cosmological observations is of paramount importance for gaining insight into both the mechanism responsible for generating the primordial perturbations (whether inflation or alternative scenarios), as well as the future evolution of the Universe. It is well known that $\Omega_K$ can be constrained from measurements of the CMB temperature anisotropy power spectrum. However, the strength of such constraints will be limited by the so-called geometrical degeneracy, first discussed in~\cite{Bond:1997wr,Zaldarriaga:1997ch,Efstathiou:1998xx}. The geometrical degeneracy is in essence reflecting the fact that an important part of the cosmological information contained in the CMB resides in the acoustic angular scale $\theta_s$. The acoustic angular scale is given by the ratio of the comoving sound horizon $r_s$ to the comoving angular diameter distance $D_A$, both evaluated at last-scattering, and controls the position of the first acoustic peak.

Let us imagine fixing $r_s$ by keeping early Universe physics unchanged, and more precisely by keeping $\Omega_bh^2$ and $\Omega_ch^2$ fixed. Then, there are various combinations of the matter density parameter $\Omega_m$, Hubble constant $H_0$, and spatial curvature parameter $\Omega_K$ which lead to the same value of $D_A$ and hence the same value of $\theta_s$, as a result keeping the CMB temperature anisotropy power spectrum to a large extent unaltered.~\footnote{This is not completely true, as changes in $\Omega_m$ which do not keep $\Omega_mh^2$ fixed (with $h=H_0/(100\,{\rm km}\,{\rm s}^{-1}\,{\rm Mpc}^{-1})$ the reduced Hubble constant) will affect both the early integrated Sachs-Wolfe effect (which affects the height of all acoustic peaks) and the amount of gravitational lensing (which smooths the higher order acoustic peaks). Moreover, changes in $\Omega_m$, $H_0$, and $\Omega_K$ will also change the dark energy density parameter $\Omega_{\Lambda}$, although the effect of the latter on the late integrated Sachs-Wolfe effect occurs exclusively on large scales which are swamped by cosmic variance.} To put it differently, along the corresponding geometrical degeneracy direction there are several combinations of the parameters $\Omega_m$-$H_0$-$\Omega_K$ which produce approximately the same CMB temperature anisotropy power spectrum as that of a spatially flat model ($\Omega_K=0$) with given values of $\Omega_m$ and $H_0$. These degeneracies act in such a way that the $\Omega_K$ posterior might be skewed towards negative values, a result known since the time of BOOMERanG~\citep[][]{Melchiorri:2000px}. With these caveats in mind, \textit{Planck} 2018 CMB temperature and polarization anisotropy data alone set the constraint $\Omega_K=-0.044^{+0.018}_{-0.015}$, which at first glance appears to suggest a spatially closed Universe~\citep{Aghanim:2018eyx}.

However, measuring the gravitational lensing-induced smoothing of the higher order acoustic peaks on small scales helps to better determine $\Omega_m$ and hence alleviate the geometrical degeneracy. In fact, the simulations of~\cite{DiValentino:2019qzk} showed that a \textit{Planck}-like experiment should be able to constrain spatial curvature to $2\%$ accuracy without introducing any significant bias towards closed models. This was also beautifully demonstrated by both the \textit{Planck} collaboration when including measurements of the CMB lensing power spectrum reconstructed from the temperature 4-point function~\citep[][]{Aghanim:2018oex}, and by the ACT collaboration~\citep[][]{Aiola:2020azj}, which in both cases find the data to be consistent with $\Omega_K=0$.

The previous discussion shows that CMB angular power spectra remain, at least in principle, the only cosmological observable which can provide important constraints on $\Omega_K$ without the addition of external datasets. Nonetheless, it remains important and desirable to improve the reliability of CMB-only constraints on $\Omega_K$ by combining CMB measurements with other external datasets which can break or at least alleviate the geometrical degeneracy. It is clear that this can be achieved by accessing orthogonal information which helps pinning down the late-time expansion rate. An example in this sense are BAO distance and expansion rate measurements, with anisotropic BAO measurements separately constraining $r_s/D_A$ and $Hr_s$. In fact, various works have discussed the combination of CMB data from \textit{Planck} with BAO measurements to break the geometrical degeneracy and provide tighter constraints on curvature. For example, combining P18 data with recent BAO measurements gives $\Omega_K=0.0008 \pm 0.0019$~\citep[see][]{Aghanim:2018eyx,Handley:2019tkm,DiValentino:2019qzk,Efstathiou:2020wem,Vagnozzi:2020zrh}, in perfect agreement with the Universe being spatially flat. Similar, although slightly less constraining results, can be obtained by combining P18 data with the full-shape (FS) galaxy power spectrum measured from the BOSS DR12 CMASS sample, with $\Omega_K=0.0023 \pm 0.0028$ obtained from this combination as shown in~\cite{Vagnozzi:2020zrh}, see also related important work in~\cite{Chudaykin:2020ghx}. Both BAO and FS data help in breaking the geometrical degeneracy and improving CMB-only constraints on $\Omega_K$, as discussed in~\cite{Efstathiou:2020wem,Chudaykin:2020ghx,Vagnozzi:2020zrh}.

However, the question of whether the dataset combinations discussed above are legitimate in first place remains. If two datasets are in tension with each other within an assumed cosmological model, the resulting parameter constraints should be viewed with caution, regardless of the ability of one dataset to break important parameter degeneracies inherently present in the other dataset. Unfortunately, this is the case with both BAO and FS data, as discussed in~\cite{Handley:2019tkm} and~\cite{DiValentino:2019qzk} in the case of BAO measurements, and~\cite{Vagnozzi:2020zrh} in the case of FS measurements: both datasets are in relatively strong tension with P18 data under the assumption of a curved Universe. In the words of~\cite{Handley:2019tkm}, ``\textit{conclusions regarding the spatial curvature of the universe which stem from the combination of these data should therefore be viewed with suspicion}''. As shown in~\cite{Handley:2019tkm,DiValentino:2019qzk,DiValentino:2020hov}, similar levels of tension are present between P18 data and: \textit{Planck} CMB lensing data; SNeIa distance moduli measurements from the \textit{Pantheon} sample; and local measurements of $H_0$.

It therefore appears as if we are standing at an impasse: we would like to improve constraints on $\Omega_K$ from P18 data by breaking the geometrical degeneracy. However, any attempt to do so must rely on the combination of P18 with datasets (CMB lensing, BAO, full-shape galaxy power spectrum, SNeIa distance moduli, local $H_0$ measurements) which are in tension with P18, an issue which raises questions as to the legitimacy of the combination. An additional somewhat minor setback is that at more than one point in the data reduction process leading to the final BAO and FS measurements, several (fiducial) cosmological model assumptions are injected. In other words, a fiducial cosmological model, in all cases $\Lambda$CDM, has to be assumed in order to go from the starting galaxy catalog to the final BAO and FS measurements. For models which are not far from $\Lambda$CDM, the associated residual model-dependence is expected to be small~\citep[see e.g.][]{Ding:2017gad,Sherwin:2018wbu,Carter:2019ulk,Bernal:2020vbb}. However, for more extreme models, this might no longer be the case, as shown for instance in~\cite{Heinesen:2019phg}. The question is then whether a cosmology with $\Omega_K$ as large as $\vert \Omega_K \vert \sim {\cal O}(10^{-2}-10^{-1})$ as suggested by \textit{Planck} temperature and polarization data is extreme enough for these fiducial cosmology assumptions to be a concern.

With these issues in mind, it becomes clear that to convincingly resolve the spatial curvature conundrum one should look for one or more datasets which can be safely combined with \textit{Planck} CMB temperature and polarization anisotropy measurements to constrain $\Omega_K$ by breaking the geometrical degeneracy while avoiding the drawbacks listed above. Therefore, such a dataset should satisfy the following characteristics, roughly in order of decreasing importance:
\begin{itemize}
    \item when combined with P18 data, it should help break the geometrical degeneracy;
    \item it should \textit{not} be in strong tension with P18 data when working within a non-flat Universe;
    \item it should contain little or no amount of (fiducial) cosmological model-dependent assumptions.
\end{itemize}
Is there a cosmological measurement satisfying all these characteristics, while still allowing for competitive constraints on $\Omega_K$ once combined with \textit{Planck}? In this work, we will argue that the answer is ``\textit{yes}'', and takes the form of cosmic chronometer measurements of $H(z)$.

\subsection{Cosmic chronometers}
\label{subsec:cc}

The principle underlying the use of cosmic chronometers (CC) to measure the Hubble parameter as a function of redshift, $H(z)$, was first proposed by one of us in~\cite{Jimenez:2001gg}, and is based upon the relation between time $t$, redshift $z$, and Hubble parameter $H(z)$ in a FLRW Universe:
\begin{eqnarray}
H(z) = -\frac{1}{1+z}\frac{dz}{dt}\,.
\label{eq:dtdz}
\end{eqnarray}
Therefore, cosmological model-independent measurements of $H(z)$ can in principle be obtained from high-fidelity measurements of $dz$ and $dt$. While redshifts can be measured to $0.1\%$ precision via spectroscopy of extragalactic objects, the main difficulty in the use of Eq.~(\ref{eq:dtdz}) is that of determining the corresponding differential age evolution $dt$, which requires a ``cosmic chronometer''.

The ideal contestant to play the role of CC is constituted by passive stellar populations which evolve on a timescale much larger than their differential ages: an example is that of massive ($\log_{10}(M/M_{\odot}) \gtrsim 11$), early, passively-evolving galaxies. In fact, various works have found that these galaxies formed and assembled their mass at high redshift ($z \sim 2$-$3$) and over a very short period of time ($t \lesssim 0.3\,{\rm Gyr}$), before quickly exhausting their gas reservoir and hence evolving passively~\citep[see for instance][]{Cimatti:2004gq,Treu:2005fk,Pozzetti:2009dm,Thomas:2009wg,Choi:2014vfa,Onodera:2014rpa,2016A&A...592A..19C,2016ApJ...832...79P,2018MNRAS.480.4379C,2019ApJ...874...17B,2019ApJ...870..133E,Moresco:2020fbm}. Measurements of the age difference $\Delta t$ between two passively-evolving galaxies which formed at the same time and are separated by a small redshift interval $\Delta z$ around $z_{\rm eff}$ can be used to estimate $dz/dt \approx \Delta z/\Delta t$, and hence $H(z_{\rm eff})$ through Eq.~(\ref{eq:dtdz}).

\begin{table}[!t]
\begin{center}
\scalebox{0.65}{\begin{tabular}{| c | c | c | c |} 
    \hline \hline
    $z$ & $H(z) \ ({\rm km}\,{\rm s}^{-1}\,{\rm Mpc}^{-1}) $ & References \\
    \hline \hline
    0.09 & 69 $\pm$ 12 & ~\cite{Jimenez:2003iv} \\
    \hline \hline
    0.17 & 83 $\pm$ 8 & ~\cite{Simon:2004tf} \\
    0.27 & 77 $\pm$ 14 & \\
    0.4 & 95 $\pm$ 17 & \\
    0.9 & 117 $\pm$ 23 & \\
    1.3 & 168 $\pm$ 17 & \\
    1.43 & 177 $\pm$ 18 & \\
    1.53 & 140 $\pm$ 14 & \\ 
    1.75 & 202 $\pm$ 40 & \\
    \hline \hline
    0.48 & 97 $\pm$ 62 & ~\cite{Stern:2009ep} \\
    0.88 & 90 $\pm$ 40 & \\
    \hline \hline
    0.1791 & 75 $\pm$ 4 & ~\cite{Moresco:2012by} \\ 
    0.1993 & 75 $\pm$ 5 & \\
    0.3519 & 83 $\pm$ 14 & \\
    0.5929 & 104 $\pm$ 13 & \\
    0.6797 & 92 $\pm$ 8 & \\
    0.7812 & 105 $\pm$ 12 & \\
    0.8754 & 125 $\pm$ 17 & \\
    1.037 & 154 $\pm$ 20 & \\
    \hline \hline   
    0.07 & 69 $\pm$ 19.6 & ~\cite{Zhang:2012mp} \\
    0.12 & 68.6 $\pm$ 26.2 & \\  
    0.2 & 72.9 $\pm$ 29.6 & \\
    0.28 & 88.8 $\pm$ 36.6 & \\ 
    \hline \hline
    1.363 & 160 $\pm$ 33.6 & ~\cite{Moresco:2015cya} \\
    1.965 & 186.5 $\pm$ 50.4 & \\ 
    \hline \hline
    0.3802 & 83 $\pm$ 13.5 & ~\cite{Moresco:2016mzx} \\
    0.4004 & 77 $\pm$ 10.2 & \\
    0.4247 & 87.1 $\pm$ 11.2 & \\
    0.4497 & 92.8 $\pm$ 12.9 & \\
    0.4783 & 80.9 $\pm$ 9.0 & \\
    \hline \hline
    0.47 & 89.0 $\pm$ 23.0 & ~\cite{Ratsimbazafy:2017vga} \\
\hline
\end{tabular}}
\caption{Latest compilation of cosmic chronometer measurements of $H(z)$, which we make use of in this work. In the first, second, and third column we report the redshift, measurement of $H(z)$, and reference for the measurement respectively.}
\label{tab:cc}
\end{center}
\end{table}

The use of CC to infer $H(z)$ in a cosmology-independent way has been the subject of much study in the last 20 years, and present CC measurements have determined $H(z)$ up to $z \approx 2$ with a typical $\lesssim 10\%$ uncertainty~\citep[see e.g.][including works by one of us]{Jimenez:2003iv,Simon:2004tf,Stern:2009ep,Moresco:2012by,Zhang:2012mp,Moresco:2015cya,Moresco:2016mzx,Ratsimbazafy:2017vga}. The latest compilation of CC measurements comprise $31$ measurements of $H(z)$ in the range $0.07<z<1.965$, which we summarize in Tab.~\ref{tab:cc} alongside the corresponding references. These measurements have been extensively used to constrain cosmological parameters, both alone or in combination with other probes, as well as within the standard $\Lambda$CDM+GR cosmological model or in alternative scenarios.~\footnote{For an inevitably incomplete list of examples of works in this direction, see e.g.~\cite{Capozziello:2014zda,Moresco:2016nqq,Nunes:2016dlj,LHuillier:2016mtc,Verde:2016ccp,Nunes:2016drj,Sola:2016hnq,Zhao:2017cud,Haridasu:2017lma,Moresco:2017hwt,Ovgun:2017iwg,Capozziello:2017buj,Pan:2017zoh,Haridasu:2018gqm,Yang:2018euj,Saridakis:2018unr,DAgostino:2019wko,Benetti:2019gmo,Krishnan:2020obg,DAgostino:2020dhv,Singirikonda:2020ieg,Capozziello:2020ctn,DiValentino:2020evt,LeviSaid:2020mbb,Anagnostopoulos:2020ctz,Luo:2020ufj,deMartino:2020yhq,Rudra:2020nxk,Aljaf:2020nsl,Odintsov:2020qzd,Bonilla:2020wbn,Renzi:2020fnx}.} The impact of observational systematics on the CC dataset, and therefore on our results, will be discussed in more detail in Appendix~\ref{sec:appendix}, and found to be small.

Going back to the three characteristics of the ideal dataset to be combined with P18 data and shed further light on the spatial curvature conundrum, outlined at the end of Section~\ref{subsec:curvaturegeometrical}, it is quite clear that measurements of $H(z)$ at late times can help enormously in alleviating the geometrical degeneracy, a fact that had already been appreciated earlier by one of us in~\cite{Moresco:2016mzx}. The reason is that $H(z)$ directly appears in the integral determining $D_A$ (and hence $\theta_s$). Furthermore, this integral picks up most of its contributions at late times, which happen to be precisely those constrained by CC. Moreover, as we already discussed, CC deliver measurements of the late-time expansion history which are virtually free of any cosmological model assumption.

In passing, we also briefly note two additional minor advantages of CC over BAO measurements in breaking the geometrical degeneracy. First of all, unlike BAO (and to a similar extent FS) measurements which require a measurement of/prior on the sound horizon (usually coming from the CMB, or a Big Bang Nucleosynthesis prior on $\omega_b$), CC do not require any external cosmological calibration whatsoever, as they directly probe the absolute scale of $H(z)$. Next, due to the integral nature of (comoving, angular diameter, or luminosity) distances in an expanding Universe, expansion history rather than distance measurements are of greater help in alleviating the geometrical degeneracy~\citep[see e.g. related earlier discussions in][]{Maor:2000jy,Maor:2001ku}. Of course, the overall advantages of CC over BAO measurements are partially offset by the larger uncertainties in the former.

The rest of this work will therefore be devoted to 1) using real data to show how CC help breaking the geometrical degeneracy and delivering more robust constraints on $\Omega_K$, and 2) checking whether CC measurements satisfy the second characteristic we outlined at the end of Section~\ref{subsec:curvaturegeometrical}, \textit{i.e.} not being in strong tension with P18 when working within a curved Universe, thus making the P18+CC combination within such a model legitimate. In the following Section, we discuss the observational datasets and analysis methods we use of in order to address these questions.

\section{Datasets and methods}
\label{sec:methods}

In this Section, we describe the cosmological observations and analysis methods we make use of in the rest of the work. Data-wise, we use two different classes of cosmological observations:
\begin{itemize}
    \item Measurements of CMB temperature and polarization anisotropies, as well as their cross-correlation, from the \textit{Planck} 2018 legacy data release~\citep[][]{Akrami:2018vks,Aghanim:2018eyx,Aghanim:2019ame}. We note that this dataset is typically referred to as \textit{Planck TTTEEE+lowE} in the papers by the \textit{Planck} collaboration, while we refer to these measurements as \textbf{\textit{Planck}}.
    \item 31 cosmic chronometer measurements of $H(z)$ in the range $0.07<z<1.965$, compiled across the years in~\cite{Jimenez:2003iv,Simon:2004tf,Stern:2009ep,Moresco:2012by,Moresco:2015cya,Moresco:2016mzx,Ratsimbazafy:2017vga}, including works by one of us. We refer to these measurements, summarized in Tab.~\ref{tab:cc}, as \textbf{\textit{CC}}.
\end{itemize}
We will therefore consider results obtained both using \textit{Planck} data alone, as well as from the \textit{Planck}+\textit{CC} dataset combination, which helps breaking the geometrical degeneracy present in CMB data alone and stabilizing the corresponding constraints on $\Omega_K$. It is important to note that we do not consider measurements of the CMB lensing power spectrum from \textit{Planck}, reconstructed from the temperature 4-point function~\citep{Aghanim:2018oex}. The reason is that we only are interested in checking whether the preference for a closed Universe from \textit{Planck} temperature and polarization anisotropies alone survives once \textit{CC} data is included, and the inclusion of lensing measurements has already been addressed in many papers~\citep[see for instance][]{Handley:2019tkm,DiValentino:2019qzk,Efstathiou:2020wem}. As shown in~\cite{Handley:2019tkm}, the \textit{Planck} CMB lensing power spectrum measured by~\cite{Aghanim:2018oex} is in $\simeq 2.5\sigma$ tension with \textit{Planck}'s CMB temperature and polarization anisotropies within the assumption of a non-flat Universe, implying that the inclusion of CMB lensing data should be done with caution.

Model-wise, we begin by considering a 7-parameter model which extends the usual 6-parameter $\Lambda$CDM model by allowing the curvature parameter $\Omega_K$ to vary. We refer to the corresponding model as the $K\Lambda$CDM model. As done by the \textit{Planck} collaboration~\citep{Aghanim:2018eyx}, we adopt an uniform prior on $\Omega_K \in [-0.3,0.3]$. However, as noted in~\cite{Efstathiou:2020wem}, an uniform prior on $\Omega_K$ might not necessarily be highly motivated from first principles (e.g. from inflation): as a result, the posterior distribution one obtains for $\Omega_K$ should not be over-interpreted.

\begin{table}[!t]
\begin{center}
\begin{tabular}{|c||c|c|c|c|}
\hline
\textbf{Model} & $\Omega_K$ & $w$ & $M_{\nu}$ & Number of parameters \\
\hline
\hline
$K\Lambda$CDM & $[-0.3,0.3]$ & $-1$ (fixed) & $0.06\,{\rm eV}$ (fixed) & 7 \\
\hline
$Kw$CDM & $[-0.3,0.3]$ & $[-3,1]$ & $0.06\,{\rm eV}$ (fixed) & 8 \\
\hline
$M_{\nu}K\Lambda$CDM & $[-0.3,0.3]$ & $-1$ (fixed) & $[0,5]\,{\rm eV}$ & 8 \\
\hline
\end{tabular}
\caption{Summary of the treatment of beyond-$\Lambda$CDM parameters (curvature parameter $\Omega_K$, dark energy equation of state $w$, and sum of the neutrino masses $M_{\nu}$) in the three cosmological models we are considering. Within square brackets we indicate the lower and upper edges of the uniform priors imposed on these parameters, if these are not fixed.}
\label{tab:models}
\end{center}
\end{table}

As recently pointed out in a number of works, including~\cite{DiValentino:2020hov}, when working within a curved Universe it is important to check the stability of the obtained parameter constraints (particularly for $\Omega_K$) against a larger parameter space. Two of the cosmological parameters most strongly correlated with $\Omega_K$ are the dark energy equation of state (DE EoS) $w$ and the sum of the neutrino masses $M_{\nu}$, respectively fixed to $w=-1$ and $M_{\nu}=0.06\,{\rm eV}$ in the $K\Lambda$CDM model. We therefore consider two one-parameter extensions of the baseline $K\Lambda$CDM model. In the first instance we also vary the DE EoS $w$, and refer to this eight-parameter model as $Kw$CDM~\citep[see for instance][for previous analyses of this kind]{Linder:2005nh,Polarski:2005jr,Huang:2006er,Clarkson:2007bc,Ichikawa:2005nb,Zhao:2006qg,Wang:2007mza,Barenboim:2009ug}. In the second instance we vary the sum of the three active neutrino masses $M_{\nu}$, and refer to this eight-parameter model as $M_{\nu}K\Lambda$CDM. We set uniform priors on $w$ and $M_{\nu}$ unless otherwise specified. The main features of these three models, and in particular the prior edges for the non-$\Lambda$CDM parameters, are summarized in Table~\ref{tab:models}.

We use the Boltzmann solver \texttt{CAMB}~\citep[][]{Lewis:1999bs} to obtain theoretical predictions for the CMB power spectra and the background expansion. We use Monte Carlo Markov Chain (MCMC) methods to sample the posterior distributions for the parameters of the three cosmological models considered, with MCMC chains generated using the cosmological sampler \texttt{CosmoMC}~\citep[][]{Lewis:2002ah}.~\footnote{The patch to \texttt{CosmoMC} to include the likelihood for the \textit{CC} measurements used is publicly available at \href{https://github.com/sunnyvagnozzi/CosmoMC-patches/tree/master/Cosmic_clocks}{github.com/sunnyvagnozzi/CosmoMC-patches/tree/master/Cosmic\_clocks}.} We monitor the convergence of the generated chains via the Gelman-Rubin parameter $R-1$~\citep[][]{Gelman:1992zz}, requiring $R-1<0.01$ for the chains to be considered converged. Finally, we compute the (logarithm of the) Bayes factor of the $K\Lambda$CDM model with respect to the $\Lambda$CDM model, in order to assess the preference for a closed Universe (if any) from a Bayesian model comparison point of view, for both the \textit{Planck} and \textit{Planck}+\textit{CC} combinations. We compute the Bayes factors of the two models directly from our chains, by making use of the \texttt{MCEvidence} code~\citep{Heavens:2017afc}.

In addition to providing constraints on cosmological parameters from the \textit{Planck}+\textit{CC} combination, we also want to assess the concordance/discordance between \textit{Planck} and \textit{CC} within a curved Universe. Various measures of concordance and discordance have been discussed in the literature~\cite[see for instance][for a selection of recent examples]{Karpenka:2014moa,MacCrann:2014wfa,Lin:2017ikq,Lin:2017bhs,Adhikari:2018wnk,Raveri:2018wln,Nicola:2018rcd,Handley:2019wlz,Handley:2019pqx,Garcia-Quintero:2019cgt,Lemos:2019txn,Raveri:2019gdp}. We will use the method first devised in~\cite{Joudaki:2016mvz,Hildebrandt:2016iqg,Joudaki:2016kym}, making use of the so-called deviance information criterion (DIC) and utilized in a similar context by~\cite{DiValentino:2019qzk} and~\cite{Vagnozzi:2020zrh}. Recall that the DIC is an information theory-based model comparison tool given by~\citep[see e.g.][]{Spiegelhalter:2002ghw}:
\begin{eqnarray}
{\rm DIC} = 2\overline{\chi^2(\theta)}-\chi^2(\hat{\theta})\,,
\label{eq:dic}
\end{eqnarray}
where $\overline{\chi^2(\theta)}$ is the average of the effective $\chi^2$ over the posterior distribution and $\chi^2(\hat{\theta})$ is the best-fit effective $\chi^2$. Let us fix the underlying cosmological model and consider two different datasets $D_1$ and $D_2$, used to place constraints on the parameters of the cosmological model. We then define the quantity ${\cal G}(D_1\,,D_2)$ by~\citep[see e.g.][]{Joudaki:2016mvz,Hildebrandt:2016iqg,Joudaki:2016kym}:
\begin{eqnarray}
{\cal G}(D_1\,,D_2) \equiv {\rm DIC}(D_1 \cup D_2)-{\rm DIC}(D_1)-{\rm DIC}(D_2)\,,
\label{eq:gd1d2}
\end{eqnarray}
where by ${\rm DIC}(D_1 \cup D_2)$ we indicate the DIC evaluated from the combination of the $D_1$ and $D_2$ datasets. We construct the quantity ${\cal I}(D_1\,,D_2)$, later used to estimate the concordance between $D_1$ and $D_2$, by~\citep[see e.g.][]{Joudaki:2016mvz,Hildebrandt:2016iqg,Joudaki:2016kym}:
\begin{eqnarray}
{\cal I}(D_1\,,D_2) \equiv \exp \left [ -\frac{{\cal G}(D_1\,,D_2)}{2} \right ] \,.
\label{eq:id1d2}
\end{eqnarray}
Finally, we can use ${\cal I}$ to estimate the level of concordance/discordance between \textit{Planck} and \textit{CC} within a curved Universe, with $\log{\cal I}>0$ [$\log{\cal I}<0$] indicating agreement [disagreement] between the two datasets within the assumed cosmological model. To qualify the level of discordance given a value of $\log{\cal I}<0$, we follow the Jeffreys-like scale introduced in this context in~\cite{DiValentino:2019qzk}, and consider the level of discordance between \textit{Planck} and \textit{CC} to be ``mild'' if $\vert \log{\cal I} \vert <0.5$,  ``definite'' if $\vert \log{\cal I} \vert >0.5$, ``strong'' if $\vert \log{\cal I} \vert >1.0$, and ``decisive'' if $\vert \log{\cal I} \vert >2.0$.

\begin{figure}[!t]
\centering
\includegraphics[width=0.7\textwidth]{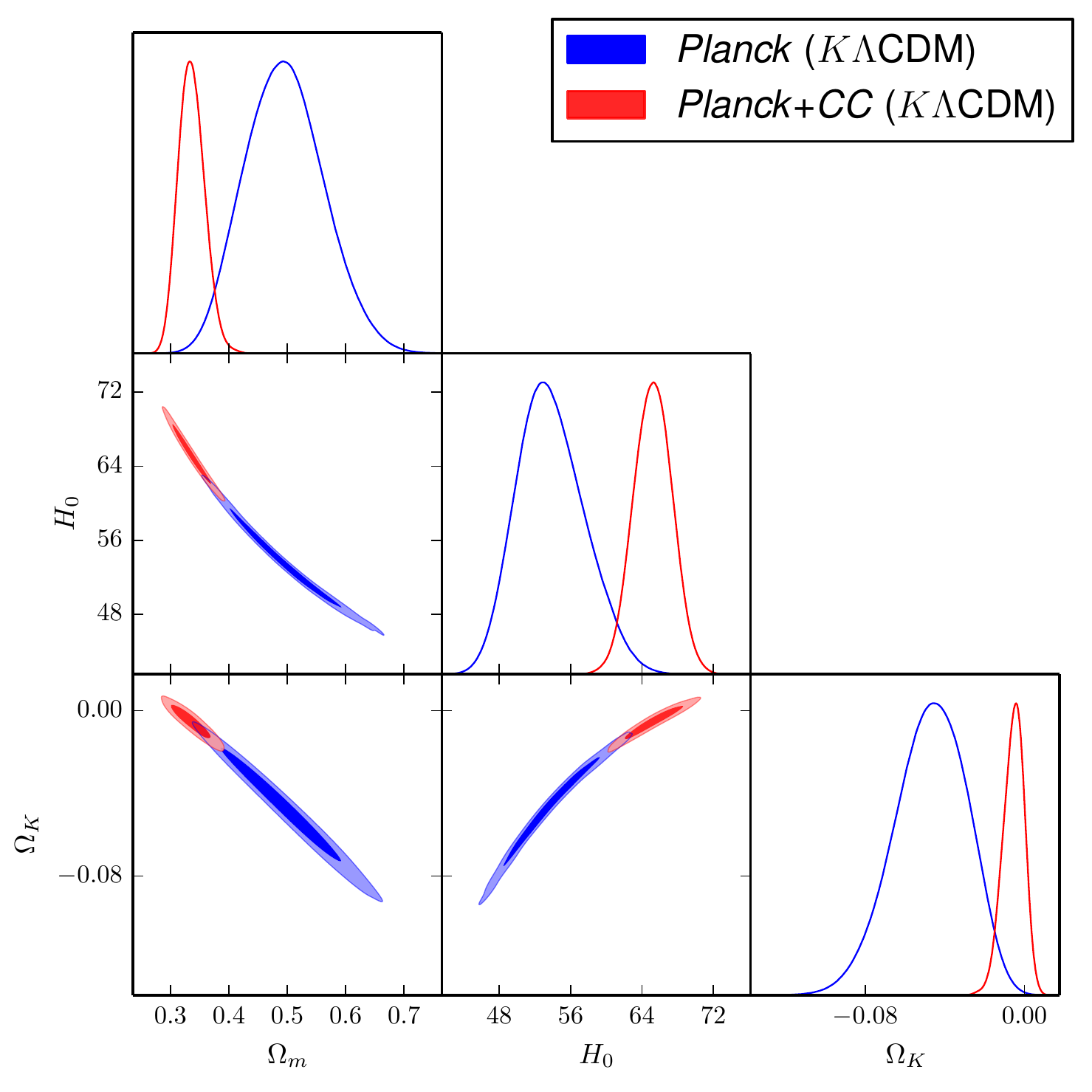}
\caption{Triangular plot showing 2D joint and 1D marginalized posterior probability distributions for $\Omega_m$, $H_0$, and $\Omega_K$ from the \textit{Planck} (blue contours) and \textit{Planck}+\textit{CC} (red contours) dataset combinations, obtained within the 7-parameter $K\Lambda$CDM model. It is clear that combining \textit{Planck} with \textit{CC} helps break the geometrical degeneracy, pushing the constraints on $\Omega_m$ and $H_0$ towards values more in line with independent late-time measurements, and pushing the constraints on $\Omega_K$ towards the spatially flat case $\Omega_K=0$.}
\label{fig:base_omegak_tri}
\end{figure}

\section{Results}
\label{sec:results}

\begin{table*}[!t]
\centering
\scalebox{1.2}{
\begin{tabular}{|c||cc|cc|}       
\hline\hline
Parameters & \multicolumn{2}{c|}{$K\Lambda$CDM} \\
 & \textit{Planck} & \textit{Planck}+\textit{CC} \\ \hline
$\Omega_K$ & $-0.044^{+0.018}_{-0.015}$ & $-0.0054 \pm 0.0055$ \\
$H_0\,[{\rm km}\,{\rm s}^{-1}\,{\rm Mpc}^{-1}]$ & $54.36^{+3.25}_{-3.96}$ & $65.23 \pm 2.14$ \\
$\Omega_m$ & $0.485^{+0.058}_{-0.068}$ & $0.336 \pm 0.022$ \\
\hline \hline                                                  
\end{tabular}}
\caption{68\%~C.L. constraints on the matter density parameter $\Omega_m$, the Hubble constant $H_0$, and the curvature density parameter $\Omega_K$, within the seven-parameter $K\Lambda$CDM model.}
\label{tab:parametersklcdm}                                              
\end{table*}

We first discuss the results obtained within the minimal 7-parameter $K\Lambda$CDM model. In Tab.~\ref{tab:parametersklcdm} we report 68\%~C.L. constraints on $\Omega_K$, $H_0$, and $\Omega_m$ from the \textit{Planck} and \textit{Planck}+\textit{CC} datasets, within the $K\Lambda$CDM model. We see from Tab.~\ref{tab:parametersklcdm} the well-known result that $\Omega_K=-0.044^{+0.018}_{-0.015}$ at 68\%~C.L. from \textit{Planck} alone within the $K\Lambda$CDM model, which corresponds to an apparent strong indication for a closed Universe. Due to the direction of the $\Omega_m$-$H_0$-$\Omega_K$ geometrical degeneracy we discussed in Section~\ref{sec:curvatureandchronometers}, the large and negative value of $\Omega_K$ is compensated by rather extreme values of $H_0=54.36^{+3.25}_{-3.96}\,{\rm km}\,{\rm s}^{-1}\,{\rm Mpc}^{-1}$ and $\Omega_m=0.485^{+0.058}_{-0.068}$, respectively much lower and much higher than those obtained within $\Lambda$CDM. Importantly, these values are also in strong tension with those obtained from independent late-time measurements. For example, the value of $H_0$ is in strong tension with a wide host of local measurements~\citep[see e.g.][]{Riess:2019cxk,Wong:2019kwg,Freedman:2019jwv,Huang:2019yhh,Pesce:2020xfe}, whereas the value of $\Omega_m$ is in strong tension with independent late-time measurements from cosmic shear and cluster counts~\citep[see e.g.][]{Ade:2015fva,Sakr:2018new,Zubeldia:2019brr,Abbott:2020knk,Asgari:2020wuj}.

Once \textit{Planck} is combined with the \textit{CC} dataset, the geometrical degeneracy is broken, and the inferred values of $\Omega_m$ and $H_0$ are much more in line with the independent late-time probes we mentioned previously. For instance, we find $H_0=65.23 \pm 2.14\,{\rm km}\,{\rm s}^{-1}\,{\rm Mpc}^{-1}$, in much better agreement with local measurements than the previous $H_0 \approx 54\,{\rm km}\,{\rm s}^{-1}\,{\rm Mpc}^{-1}$ (barring of course the long-standing Hubble tension). Importantly, we find $\Omega_K=-0.0054 \pm 0.0055$, consistent with a spatially flat Universe within $1\sigma$, also in line with results obtained using other late-time probes to break the geometrical degeneracy (such as BAO and FS galaxy power spectrum measurements). Recall, for instance, that combining \textit{Planck} with BAO measurements gives $\Omega_K=0.0008 \pm 0.0019$~\citep{Aghanim:2018eyx,Efstathiou:2020wem}, while combining \textit{Planck} with FS galaxy power spectrum measurements from the BOSS DR12 CMASS sample gives $\Omega_K=0.0023 \pm 0.0028$~\citep{Vagnozzi:2020zrh}. From these numbers one important point we note is that, despite the overall larger uncertainties in the \textit{CC} dataset, the sensitivity to $\Omega_K$ once combined with \textit{Planck} data is comparable to that obtained combining \textit{Planck} with BAO or FS data. Compared to the case where \textit{Planck} data is combined with BAO (FS) measurements, the uncertainty on $\Omega_K$ is only a factor of $\approx 3$ ($\approx 2$) worse: in this context, \textit{CC} data is therefore competitive with these more widely used probes. Similar considerations would hold if we used CMB lensing, SNeIa distance moduli, or local measurements of $H_0$ instead.

It is also instructive to inspect the $\Omega_m$-$H_0$-$\Omega_K$ triangular plot to gain more insight into the role of the \textit{CC} dataset in breaking the geometrical degeneracy and stabilizing constraints on $\Omega_K$. This triangular plot is shown in Fig.~\ref{fig:base_omegak_tri}, with blue and red contours corresponding to the \textit{Planck} and \textit{Planck}+\textit{CC} dataset combinations respectively. The three 2D contours in Fig.~\ref{fig:base_omegak_tri}, especially the two lowermost ones involving $\Omega_K$, highlight the crucial role played by the \textit{CC} dataset in breaking the geometrical degeneracy: recall this was one of the key aspects of the dataset we were searching for in order to shed light on the spatial curvature conundrum, and is something that had already been appreciated earlier by one of us in~\cite{Moresco:2016mzx}. The impact of observational systematics on these results is discussed in more detail in Appendix~\ref{sec:appendix} and found to be small.

\begin{figure}[!t]
\centering
\includegraphics[width=0.7\textwidth]{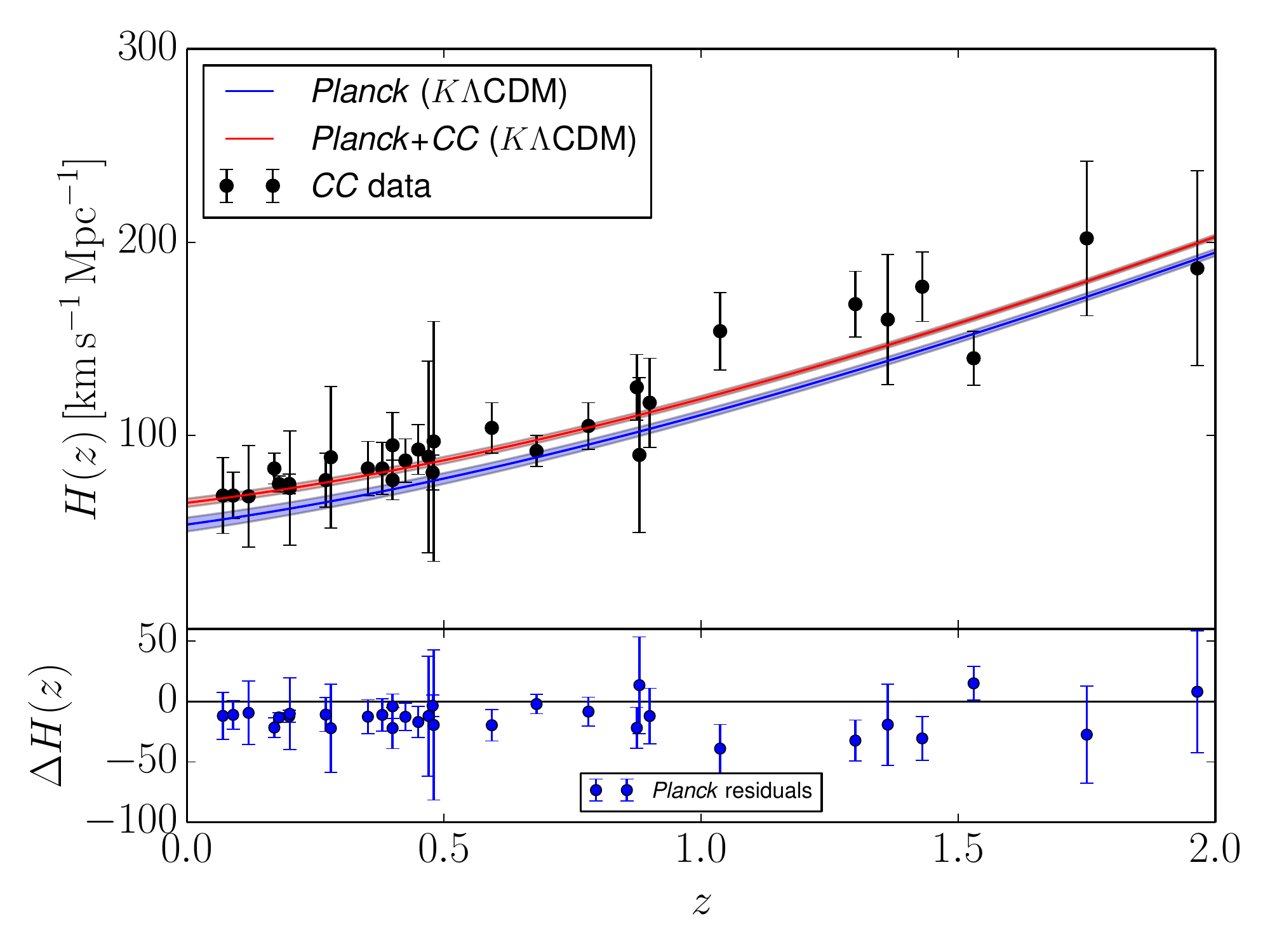}
\caption{\textit{Upper panel}: cosmic chronometer measurements correspond to the black datapoints. The two colored bands correspond to the $1\sigma$ confidence ranges for the expansion rate $H(z)$ extrapolated from a fit to the $K\Lambda$CDM model of the \textit{Planck} (blue band) and \textit{Planck}+\textit{CC} (red band) dataset combinations. \textit{Lower panel}: residuals with respect to the $K\Lambda$CDM model with cosmological parameters inferred from \textit{Planck} alone. From here it is visually clear that this set of parameters overall underpredicts $H(z)$ compared to the \textit{CC} data, especially in the first bins with the highest signal-to-noise.}
\label{fig:h_of_z}
\end{figure}

The other crucial take-away message from Fig.~\ref{fig:base_omegak_tri} is that it is visually clear that \textit{Planck} and \textit{CC} are not in strong tension \textit{even when working within the assumption of a curved Universe}, even though the agreement between them is admittedly not perfect. Nonetheless, due to the absence of strong tensions, the \textit{Planck} and \textit{CC} datasets can be safely combined even within the $K\Lambda$CDM model. This is quite unlike the case of BAO or FS galaxy power spectrum measurements, which are instead in very strong tension with \textit{Planck} when allowing spatial curvature to vary, which makes the resulting dataset combination at the very least questionable (even though we wish to stress again that these measurements are crucial in order to break the geometrical degeneracy). On this matter, we invite the reader to compare the present Fig.~\ref{fig:base_omegak_tri} to Fig.~1 in~\cite{Vagnozzi:2020zrh}: in the latter, it is visually clear that the corresponding contours are widely disjoint for both the cases where \textit{Planck} is combined with BAO or FS galaxy power spectrum measurements.

Finally, we compute the natural logarithm of the Bayes factor of the $K\Lambda$CDM model with respect to the $\Lambda$CDM model, $\ln B_{ij}$. From \textit{Planck} data alone, we find $\ln B_{ij}=+2.5$, whereas from the \textit{Planck}+\textit{CC} dataset combination we find $\ln B_{ij}=-3.4$. On the widely adopted Jeffreys scale~\citep{Jeffreys:1939xee}, these numbers correspond to a definite preference for $K\Lambda$CDM from \textit{Planck} alone (in line with previous findings in~\cite{Handley:2019tkm} and ~\cite{DiValentino:2019qzk}), and to a strong preference for $\Lambda$CDM from \textit{Planck}+\textit{CC}. Therefore, the addition of \textit{CC} data to \textit{Planck} both pushes the $\Omega_K$ constraints towards $\Omega_K=0$, and leads to $\Lambda$CDM being preferred from Bayesian model comparison considerations. We find comparable figures when using the Savage-Dickey density ratio (SDDR) instead of \texttt{MCEvidence} to compute these Bayes factors: the SDDR (first introduced in the context of cosmology in~\cite{Trotta:2005ar}) can be used in this context since the $\Omega_K$ prior is separable from the priors on the $\Lambda$CDM parameters, and the $\Lambda$CDM model is nested within the $K\Lambda$CDM model.

\subsection{Tension between \textit{Planck} and cosmic chronometers data}
\label{subsec:tension}

To further quantify the concordance or discordance between \textit{Planck} and \textit{CC} within the $K\Lambda$CDM model, we use the ${\cal I}$ diagnostic defined in Section~\ref{sec:methods}. Doing so we find $\log{\cal I}$(\textit{Planck},\textit{CC})$\approx -0.47$. Since $\log{\cal I}<0$, this indicates that the two datasets are not exactly in agreement. On the Jeffreys-like scale we are using, this corresponds to a mild disagreement between \textit{Planck} and \textit{CC}.

The above result is clear from Fig.~\ref{fig:base_omegak_tri}: while the $2\sigma$ regions in the relevant 2D contours for \textit{Planck} and \textit{Planck}+\textit{CC} overlap for all the three parameter combinations shown, the $1\sigma$ regions do not, preventing the concordance from being high. On the other hand, there is clearly no strong discordance either. The qualification of the discordance between \textit{Planck} and \textit{CC} within the $K\Lambda$CDM model being only mild correctly captures the qualitative features we visually see in Fig.~\ref{fig:base_omegak_tri}.

To further elucidate the mild disagreement between \textit{Planck} and \textit{CC} within the $K\Lambda$CDM model, in the upper panel of Fig.~\ref{fig:h_of_z} we plot the \textit{CC} dataset along with the predicted $H(z)$ extrapolated from the parameters inferred within the $K\Lambda$CDM model from the \textit{Planck} (blue band) and \textit{Planck}+\textit{CC} (red band) dataset combinations. From the upper panel of Fig.~\ref{fig:h_of_z}, we see by eye that the cosmological parameters obtained within the $K\Lambda$CDM model from \textit{Planck} alone (including the rather extreme values of both $H_0$ and $\Omega_m$, in tension with independent late-time probes as we discussed previously) appear to underpredict $H(z)$, especially in the first few and most precise redshift bins. This is clearer from the bottom panel of Fig.~\ref{fig:h_of_z}, where we plot the \textit{CC} dataset residuals with respect to this model. For the cosmological parameters obtained within the $K\Lambda$CDM model from \textit{Planck} alone, it is clear that the resulting expansion rate $H(z)$ is lower than that indicated by the \textit{CC} dataset, with most of the first redshift bins being consistently off by $\approx 1\sigma$ or more.

\subsection{Ages of the oldest objects in the Universe}
\label{subsec:ageofoldest}

The age of the Universe is an important, albeit often ignored, piece in the cosmic concordance puzzle. Within both the $\Lambda$CDM and $K\Lambda$CDM models, as well as extensions thereof, the age of the Universe $t_U$ is a \textit{prediction} given the values of the cosmological parameters inferred from CMB data. On the other hand, one can also measure or set lower limits on $t_U$ by measuring the ages of the oldest objects in the Universe. In the nineties, it was the determination of the ages of the oldest objects in the Universe which suggested that the then dominant cosmological model, the Einstein-de Sitter Universe, needed significant revision~\citep[see e.g.][]{Ostriker:1995su,Jimenez:1996at,Spinrad:1997md}, while hints for objects older than the CMB-inferred $t_U$ have sporadically reappeared afterwards~\citep[see e.g.][]{Bond:2013jka,2018IAUS..334...11C}. Subsequently, the \textit{absolute} ages of distant objects (or the lookback times thereto) were used to constrain cosmological parameters and dark energy models, through an approach which is complementary to the cosmic chronometers relative ages one~\citep[see for instance][]{Alcaniz:1999kr,Lima:2000jc,Capozziello:2004jy,Jain:2005gu,Dantas:2009vs,Samushia:2009px,Dantas:2010zh,Bengaly:2013afa,Wei:2015cva,Rana:2016gha}.

\begin{figure}[!t]
\centering
\includegraphics[width=0.7\textwidth]{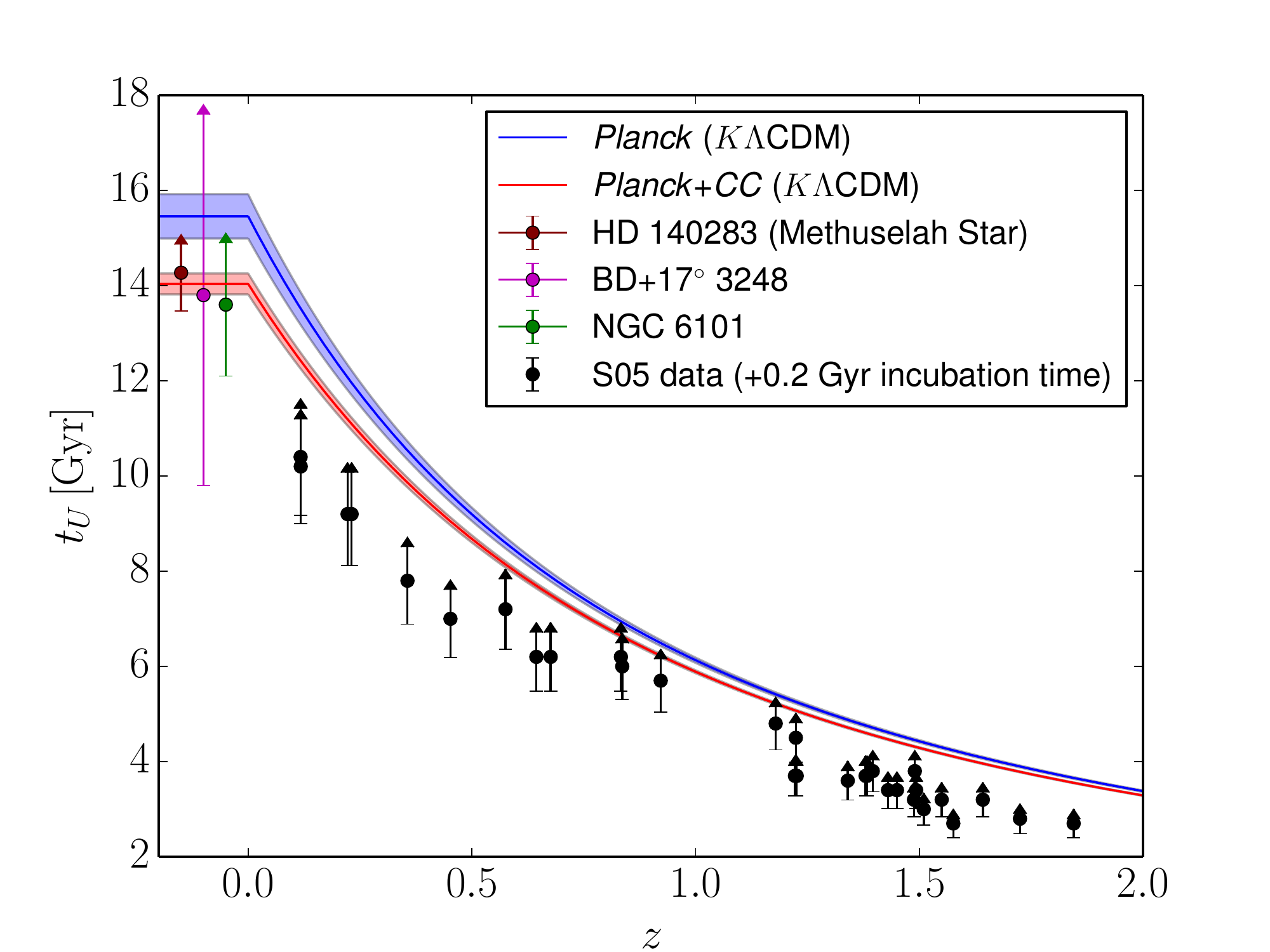}
\caption{Age-redshift relationship $t_U(z)$ extrapolated from a fit to the $K\Lambda$CDM model of the \textit{Planck} (blue band) and \textit{Planck}+\textit{CC} (red band) dataset combinations. In addition, we plot the ages of the some of the oldest objects in the Universe. The maroon, magenta, and green data points correspond to the age of the Population II halo subgiant HD 140283 (also known as Methuselah star) as determined by~\cite{2014ApJ...792..110V}, the age of the Population II star BD+$17^{\circ}\,3248$ as determined by~\cite{Cowan:2002mi}, and the age of the globular cluster NGC 6101 as determined by~\cite{2017ApJ...838..162O}. For convenience, the $x$ axis has been extended to $z<0$ to accommodate these three $z \approx 0$ objects. The black data points correspond to what we refer to as the S05 dataset, which contains the ages of 32 passively evolving galaxies, which are among the oldest objects at $0 < z \lesssim 2$, as compiled in~\cite{Simon:2004tf} and reported in Table~1 of~\cite{Samushia:2009px}. To the ages of these objects we have added an incubation time of $\approx 0.2\,{\rm Gyr}$, indicative of the typical time elapsed between the Big Bang and the time when these objects started forming, following~\cite{Jimenez:2019onw}.~ The arrows pointing upwards indicate that the ages of these objects set lower limits to the age of the Universe at the redshift in question.}
\label{fig:age_z}
\end{figure}

The importance of the ages of distant objects in the context of the ongoing Hubble and spatial curvature tensions was recently reaffirmed in~\cite{Jimenez:2019onw,Valcin:2020vav,DiValentino:2019qzk,DiValentino:2020srs}. For instance, in~\cite{DiValentino:2019qzk} it was noticed that the unrealistically low value of $H_0$ recovered from \textit{Planck} data alone within the $K\Lambda$CDM model leads to an older Universe and improves the compatibility with the ages of the oldest Population II stars, with the higher value of $H_0$ suggested within the same model by a combination of Big Bang Nucleosynthesis, BAO, and uncalibrated SNeIa data being in tension with these ages. Similarly, one can envisage different cosmologies with identical $t_U$, but very different age-redshift relationship, which can be constrained through the ages of old high-$z$ objects. Here, we will therefore qualitatively explore our results in light of the ages of the oldest objects in the Universe, both at $z=0$ and $z>0$.

One of the oldest known stellar objects at $z=0$ is the Milky Way Population II halo subgiant HD 140283, also known as Methuselah star, with an estimated age of $\approx 14.27 \pm 0.80\,{\rm Gyr}$~\citep{2014ApJ...792..110V}.~\footnote{A very recent study using \textit{Gaia} parallaxes in place of the older \textit{HST} ones led to a revision of the age of HD 140283 being $\approx 13.5 \pm 0.7\,{\rm Gyr}$~\citep{Jimenez:2019onw}.} The age of HD 140283 is still compatible with $t_U = (13.800 \pm 0.024)\,{\rm Gyr}$ as inferred by \textit{Planck} data within the $\Lambda$CDM model, although it is slightly higher than the latter. Another comparably old star is the neutron-capture enhanced ultra-metal-poor Population II star BD+$17^{\circ}\,3248$, with an estimated age of $\approx 13.8 \pm 4.0\,{\rm Gyr}$~\citep{Cowan:2002mi}. Finally, the oldest known globular cluster is NGC 6101, also known as C107, with an age of $\approx 13.6 \pm 1.5\,{\rm Gyr}$ as estimated from Monte Carlo main-sequente fitting~\citep{2017ApJ...838..162O}. More recent studies of \textit{Gaia} stars (either benchmark stars or the full DR2 sample) have found stars whose isochrone ages exceed $16\,{\rm Gyr}$, a figure which would nominally spell trouble for the $\Lambda$CDM model~\citep{2018MNRAS.481.4093S,2019MNRAS.482..895S}, although warnings against indiscriminate isochrone fitting were raised in~\cite{2019A&A...622A..27H}.

\begin{figure}[!t]
\centering
\includegraphics[width=0.5\textwidth]{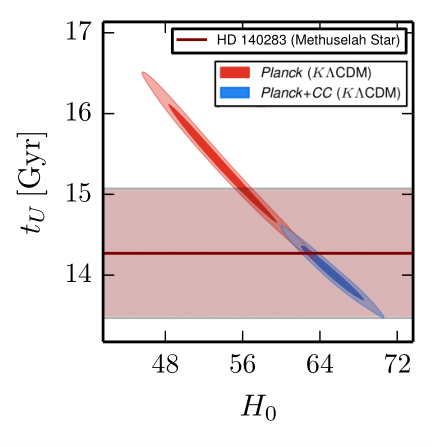}
\caption{2D joint probability distributions for the age of the Universe $t_U$ and the Hubble constant $H_0$ inferred within the $K\Lambda$CDM model from the \textit{Planck} (blue contours) and \textit{Planck}+\textit{CC} (red contours) dataset combinations. The maroon band corresponds to the $1\sigma$ confidence region for the age of the Population II halo subgiant HD 140283 (also known as Methuselah star) as determined by~\cite{2014ApJ...792..110V}.}
\label{fig:planck_cc_omegak_age_H0_new}
\end{figure}

To relate the ages of these objects to our results, in Fig.~\ref{fig:age_z} we plot the age-redshift relationship $t_U(z)$ determined from the cosmological parameters inferred within the $K\Lambda$CDM model from the \textit{Planck} (blue band) and \textit{Planck}+\textit{CC} (red band) dataset combinations. In the same Figure, we plot the ages of HD 140283, BD+$17^{\circ}\,3248$, and NGC 6101, which set lower limits on $t_U(z=0)$. We see that the lower limits on $t_U(z=0)$ set by these objects are compatible with the age of the Universe as inferred within the $K\Lambda$CDM model from the \textit{Planck}+\textit{CC} dataset combination: $t_U=(14.03 \pm 0.23)\,{\rm Gyr}$. The unrealistically low value of $H_0$ inferred within the same model from \textit{Planck} data alone leads instead to $t_U=(15.38 \pm 0.49)\,{\rm Gyr}$, which of course brings better compatibility with these ages, although we stress once more that the corresponding values of $H_0$ and $\Omega_m$ are in strong tension with those inferred from independent late-time measurements.

In addition, we also consider the ages of some of the oldest objects at $0 < z \lesssim 2$. In particular, we consider the absolute ages of 32 passively evolving galaxies as compiled in~\cite{Simon:2004tf} and reported in Table~1 of~\cite{Samushia:2009px}: we shall refer to this dataset as S05 hereafter.~\footnote{Note that these same galaxies were used to obtain the cosmic chronometer measurements of~\cite{Simon:2004tf}. These absolute ages of the S05 dataset were not reported in~\cite{Simon:2004tf}, which focused on relative ages, but were provided to the authors of~\cite{Samushia:2009px} by Raul Jim\'{e}nez via private communication, and were subsequently used to constrain cosmological parameters in several works.} We add an incubation time of $\Delta t \approx 0.2\,{\rm Gyr}$, reflecting the time elapsed between the Big Bang and the time when the objects in the S05 dataset started forming. In~\cite{Jimenez:2019onw} this was argued to be a representative number for the incubation time of the oldest and most passively-evolving galaxies, with the full distribution for $\Delta t$ given in the right panel of Fig.~1 of~\cite{Jimenez:2019onw}. This distribution is peaked at $\Delta t \approx 0.2\,{\rm Gyr}$ (consistently with earlier findings in e.g.~\cite{Samushia:2009px,Wei:2015cva}) and is slightly non-Gaussian, with tails extending as far down as $\approx 0.06\,{\rm Gyr}$ and as far up as $\approx 0.5\,{\rm Gyr}$. Within this range and for the purpose of Fig.~\ref{fig:age_z}, the exact value of the incubation time (which varies from galaxy to galaxy) is not important, and taking the representative value of $\Delta t \approx 0.2\,{\rm Gyr}$ is sufficient.~\footnote{The value $0.2\,{\rm Gyr}$ is approximately the age of the Universe at $z \sim 20$, when the first generation of low-mass stars could start to form efficiently. The reason is that only at this point could halos of virial temperatures above $10000\,{\rm K}$ form. In these halos the gas cooled by atomic hydrogen transitions and fragmented into long-lived stars.} We see from Fig.~\ref{fig:age_z} that the age-redshift relation inferred within the $K\Lambda$CDM model from the \textit{Planck}+\textit{CC} dataset combination is fully consistent with the lower limits on $t_U(z)$ set at $z>0$ by these objects.

In Fig.~\ref{fig:planck_cc_omegak_age_H0_new} we instead show 2D joint posterior probability distributions for the present age of the Universe $t_U$ and the Hubble constant $H_0$, inferred within the $K\Lambda$CDM model from both the \textit{Planck} and \textit{Planck}+\textit{CC} dataset combinations. The maroon band we plot corresponds to the $1\sigma$ confidence region for the age of HD 140283 as inferred by~\cite{2014ApJ...792..110V}. The figure confirms once more the agreement between the lower limit on $t_U$ set by HD 140283, and the value of $t_U$ inferred from \textit{Planck}+\textit{CC} within a non-flat Universe.

At present, the uncertainties in the determinations of the ages of the oldest stellar objects is dominated by the disagreement between different stellar models. These uncertainties are too large to have a decisive role in shedding further light on the cosmic concordance issues we highlighted, both in relation to the Hubble tension and the determination of spatial curvature. However, in the future we can expect substantial improvements in stellar models, which will keep improving these uncertainties considerably, consequently opening an important new window onto the age of the Universe and the previously mentioned tensions~\citep[see e.g. discussions in][]{Jimenez:2019onw}.

\subsection{Extended parameter spaces}
\label{subsec:extended}

As we discussed in Section~\ref{sec:methods}, it is important to assess the stability of constraints on spatial curvature against extensions to a larger parameter space. Therefore, we now consider two one-parameter extensions of the $K\Lambda$CDM model: the $Kw$CDM and $M_{\nu}K\Lambda$CDM models (see Tab.~\ref{tab:models}). Constraints on $\Omega_K$, $H_0$, $\Omega_m$, $w$, and $M_{\nu}$ within these models are reported in Tab.~\ref{tab:parametersextended}, for the $K\Lambda$CDM model (first column, where of course $w$ and $M_{\nu}$ are fixed to their standard values), the $Kw$CDM model (second column), and the $M_{\nu}K\Lambda$CDM model (third column).

\begin{table*}[!b]
\begin{center}                              
\scalebox{1.0}{
\begin{tabular}{|c|ccc|}       
\hline\hline
Parameter & $K\Lambda$CDM & $Kw$CDM & $M_{\nu}K\Lambda$CDM \\ \hline
$\Omega_k$ & $-0.0054 \pm 0.0055$ & $-0.0071 \pm 0.0042$ & $-0.0053 \pm 0.0056$ \\
$H_0\,[{\rm km}\,{\rm s}^{-1}\,{\rm Mpc}^{-1}]$ & $65.23 \pm 2.14$ & $72.43 \pm 5.16$ & $65.27 \pm 2.17$ \\
$\Omega_m$ & $0.336 \pm 0.022$ & $0.276 \pm 0.039$ & $0.336 \pm 0.023$ \\
$M_{\nu}\,[{\rm eV}]$ & $0.06$ (fixed) & $0.06$ (fixed) & $<0.18$ \\
$w$ & $-1$ (fixed) & $-1.32 \pm 0.21$ & $-1$ (fixed) \\
\hline\hline                                                  
\end{tabular}}
\end{center}
\caption{68\%~C.L. constraints on selected cosmological parameters ($\Omega_k$, $H_0$, $\Omega_m$, $w$, and $M_{\nu}$) within the extended eight-parameter $Kw$CDM and $M_{\nu}K\Lambda$CDM models, compared against the constraints obtained within the seven-parameter $K\Lambda$CDM model, see Table~\ref{tab:models} for descriptions of these models. All constraints have been obtained from the \textit{Planck}+\textit{CC} dataset combination. The upper limit quoted on $M_{\nu}$ is a 95\%~C.L. upper limit.}
\label{tab:parametersextended}                                              
\end{table*}

The most important result, which we read off the first row of Table~\ref{tab:parametersextended}, is that the indication for a spatially flat Universe coming from the \textit{Planck}+\textit{CC} dataset combination is relatively stable against these 1-parameter extensions. Within both extensions, the \textit{Planck}+\textit{CC} dataset combination is consistent with $\Omega_K=0$ within about $\lesssim 1.7 \sigma$. This result further reinforces the message of this paper that breaking the geometrical degeneracy by combining \textit{Planck} data with \textit{CC} data pushes towards a spatially flat Universe.

\begin{figure}[!t]
\centering
\includegraphics[width=0.7\textwidth]{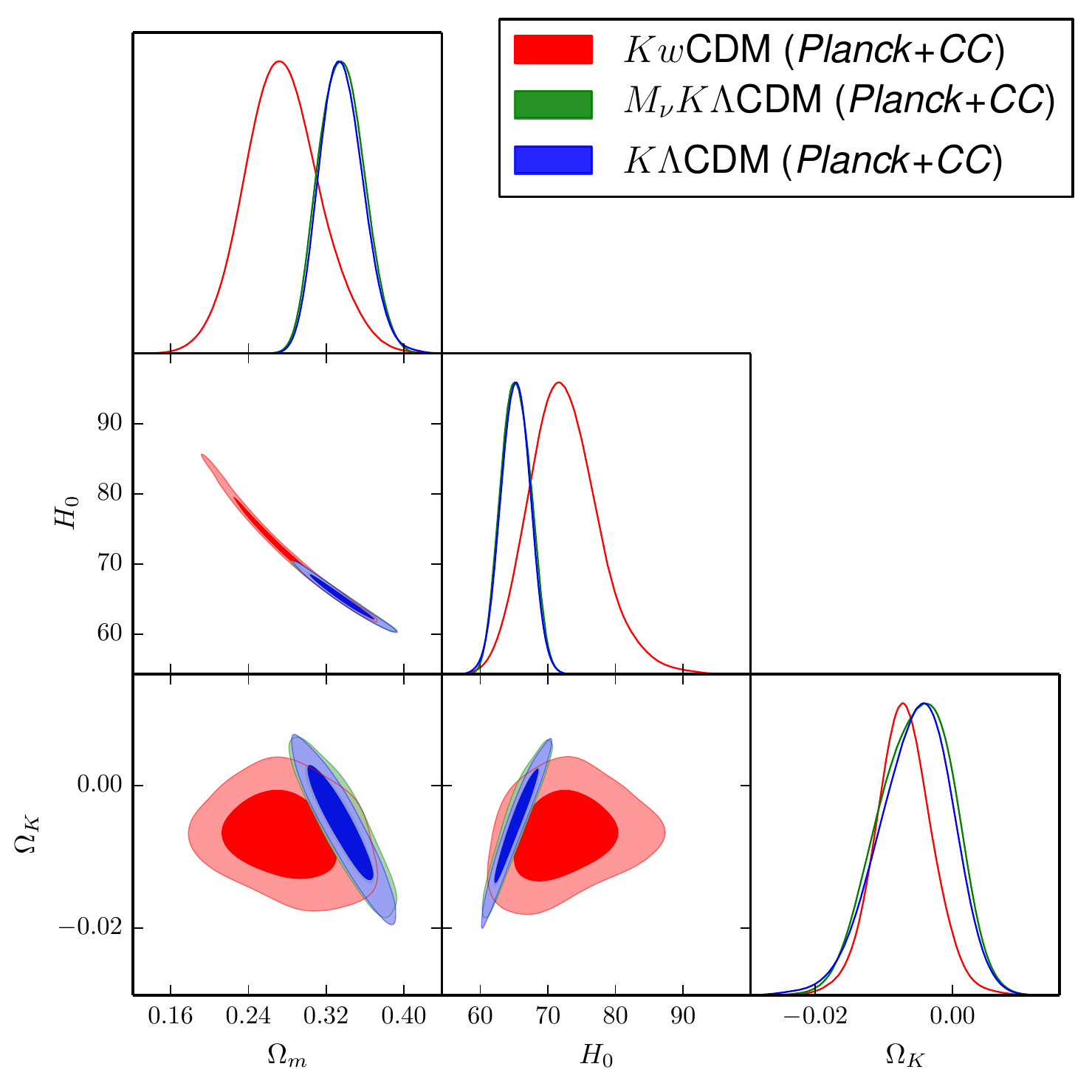}
\caption{Triangular plot showing 2D joint and 1D marginalized posterior probability distributions for $\Omega_m$, $H_0$, and $\Omega_K$ from the \textit{Planck}+\textit{CC} dataset combination, obtained within the 7-parameter $K\Lambda$CDM model (blue contours), the 8-parameter $Kw$CDM model (red contours) where the dark energy equation of state is allowed to vary, and the 8-parameter $M_{\nu}K\Lambda$CDM model (green contours) where the sum of the neutrino masses is allowed to vary. It is clear that overall the constraints on $\Omega_K$ are relatively stable against these extensions, despite the slight pull towards $\Omega_K<0$ within the $Kw$CDM model.}
\label{fig:compare_triangulars_extended}
\end{figure}

A visual representation of our results is given in the triangular plot in Fig.~\ref{fig:compare_triangulars_extended}, where we show constraints on $\Omega_m$, $H_0$, and $\Omega_K$ from the \textit{Planck}+\textit{CC} dataset combination within the $K\Lambda$CDM (blue), $M_{\nu}K\Lambda$CDM (green), and $Kw$CDM (red) models. In particular, we note from both Fig.~\ref{fig:compare_triangulars_extended} (compare the blue and green contours) as well as Tab.~\ref{tab:parametersextended} that the constraints on $\Omega_K$ are \textit{extremely} stable against the introduction of $M_{\nu}$ as an additional parameter: the differences with respect to the $K\Lambda$CDM case are hardly discernible by eye.~\footnote{Of course, the degeneracy between $M_{\nu}$ and $\Omega_K$ loosens the upper limit on $M_{\nu}$ with respect to the upper limit one would obtain within the $\Lambda$CDM+$M_{\nu}$ model: see for instance~\cite{Cuesta:2015iho,Huang:2015wrx,Giusarma:2016phn,Vagnozzi:2017ovm,Yang:2017amu,Doux:2017tsv,Upadhye:2017hdl,Nunes:2017xon,Zennaro:2017qnp,Vagnozzi:2018jhn,Giusarma:2018jei,Choudhury:2018byy,Choudhury:2018adz,Loureiro:2018pdz,Bolliet:2019zuz,RoyChoudhury:2019hls,Ivanov:2019hqk,Nunes:2020hzy,Philcox:2020vvt,Yang:2020tax} for examples of similar constraints.}

On the other hand, the results obtained within the $Kw$CDM model (second column of Tab.~\ref{tab:parametersextended}) are worthy of further comment. First of all, allowing $w$ to vary has substantially enlarged the uncertainties on both $\Omega_m$ and $H_0$ by about a factor of $2$ (while the uncertainty on $\Omega_K$ has actually slightly decreased). This is a direct consequence of the geometrical degeneracy~\citep{Bond:1997wr,Zaldarriaga:1997ch,Efstathiou:1998xx}, which was previously partially lifted by combining \textit{Planck} and \textit{CC}, but is now once more opened up more strongly due to $w$ varying.

An important effect of having allowed $w$ to vary is that the central value of $\Omega_K$ has moved towards more negative values, slightly increasing the preference for a spatially closed Universe, which however remains very weak: $\Omega_K=0$ is allowed within $\approx 1.7\sigma$. Where is this weak preference coming from? Looking at Tab.~\ref{tab:parametersextended}, we see that part of the origin is the $\approx 1.5\sigma$ indication for a phantom DE component ($w<-1$), with \textit{Planck}+\textit{CC} indicating $w=-1.32 \pm 0.21$. This preference is mostly driven by \textit{Planck} data, for which within a minimal 7-parameter $\Lambda$CDM+$w$ model one finds $w=-1.58^{+0.16}_{-0.35}$~\citep{Aghanim:2018eyx}, an apparent detection of phantom dark energy at almost $4\sigma$, but at the same time a constraint which is strongly limited by the geometrical degeneracy. It is interesting that \textit{CC} alone also provides a very weak hint for phantom DE, although with much larger error bars. Within the same minimal 7-parameter $\Lambda$CDM+$w$ model, we find $w=-1.27^{+0.80}_{-0.38}$ from \textit{CC} data alone. Combined with the stronger hint for phantom DE from \textit{Planck} alone, this leads to the weak preference for phantom DE and a spatially closed Universe once \textit{Planck} and \textit{CC} data are combined together within the $Kw$CDM model, due to the direction of the $w$-$\Omega_K$ degeneracy.

In closing, we note that the weak preference for a phantom closed Universe we have observed within the $Kw$CDM model from \textit{Planck}+\textit{CC} data has also been found in~\cite{DiValentino:2020hov} when combining \textit{Planck} with luminosity distance data from either the \textit{Pantheon} dataset (in the form of SNeIa distance moduli), or local measurements of $H_0$ from Cepheid or Tip of the Red Giant Branch-calibrated SNeIa, while allowing both $\Omega_K$ and $w$ to vary (although in that case several other parameters, up to 12 at a time, were varied as well). It is noteworthy that for the \textit{Planck}+\textit{Pantheon} dataset combination, the inferred values of $\Omega_K$ and $w$ are not far from those we inferred from the \textit{Planck}+\textit{CC} dataset combination within the $Kw$CDM model. Is this merely a coincidence or is there more to this phantom closed model? It is worth pointing out that any preference for a phantom closed model disappears when combining \textit{Planck} with BAO data, although once more this combination should be taken with caution. While we wait for more data to settle this issue, an interesting test of a possible phantom closed model has recently been proposed in~\cite{Shirokov:2020dwl}.

\section{Conclusions}
\label{sec:conclusions}

The question revolving around the spatial geometry of the Universe has been the subject of much discussion, particularly following the recent \textit{Planck} results, whose measurements of anisotropies in temperature and polarization (P18) appear at face value to prefer a spatially closed Universe with curvature parameter $\Omega_K<0$~\citep{Aghanim:2018oex}. An important point here is that in order to stabilize P18's constraints on $\Omega_K$ it is crucial to break the geometrical degeneracy present in P18 data by combining the latter with additional measurements: CMB lensing power spectrum, BAO measurements, FS galaxy power spectrum measurements, SNeIa distance moduli, and local measurements of $H_0$ just to mention a few. Combining P18 data with these probes pushes the inferred value of $\Omega_K$ towards the spatially flat case $\Omega_K=0$. The key issue, however, is that P18 is in tension with each and every single one of these probes once the assumption of a spatially flat Universe is abandoned~\citep{Handley:2019tkm,DiValentino:2019qzk,Vagnozzi:2020zrh}: because of this tension, their combination with P18 and the reliability of the resulting constraints on $\Omega_K$ should be considered with significant caution. The importance of not underestimating these tensions has recently been emphasized in a few works, including~\cite{Handley:2019tkm} and~\cite{DiValentino:2019qzk}.

In this work, our focus has been on finding a new way out of this impasse by identifying a dataset which can still reliably break the geometrical degeneracy and deliver competitive constraints on $\Omega_K$ once combined with P18 data, while not being in tension with the latter within a non-flat Universe. We have argued that cosmic chronometers (CC), \textit{i.e.} measurements of the expansion rate $H(z)$ from the relative ages of passively evolving galaxies, as first proposed by one of us in~\cite{Jimenez:2001gg}, are precisely the dataset we are looking for. Compared to some of the probes we mentioned previously, CC come with the additional extremely important advantage of being virtually free of any cosmological model assumption.

Combining P18 and CC data, we measure $\Omega_K=-0.0054 \pm 0.0055$, consistent within $1\sigma$ with the Universe being spatially flat. This result is also consistent and competitive with constraints obtained combining P18 with BAO or FS data to break the geometrical degeneracy. Importantly, we have found no substantial tension between P18 and CC within a non-flat Universe, a result which is also visually clear from Fig.~\ref{fig:base_omegak_tri}. This means that the P18+CC combination can be considered robust.

Finally, we have assessed the stability of our results against extended parameter spaces, finding them to be extremely stable against an extension where we vary the sum of the neutrino masses $M_{\nu}$. Our results are also relatively stable against an extension where we vary the dark energy equation of state $w$, with the spatially flat case $\Omega_K=0$ remaining consistent to about $1.7\sigma$, although this extension intriguingly pushes both $\Omega_K$ and $w$ towards more negative values, in the direction of a phantom closed Universe.

In conclusion, we believe our analysis represents an important step towards settling the ongoing spatial curvature debate. We have identified cosmic chronometers as a way of reliably stabilizing \textit{Planck} temperature and polarization data constraints on $\Omega_K$ by breaking the geometrical degeneracy inherent to \textit{Planck} alone while not incurring in tensions with the latter, delivering constraints competitive with those obtained combining \textit{Planck} with BAO and FS data, and carrying virtually no cosmological model assumption. Our results allow us to assert with more confidence than previous works that the Universe is indeed spatially flat to the ${\cal O}(10^{-2})$ level, something which we believe was not really possible previously due to the tensions we discussed earlier. This important result lends even more support to the already very successful inflationary paradigm, disfavoring models of incomplete inflation, with a number of e-foldings $N \approx 60$~\citep[see e.g.][for examples of such models]{Hawking:1998bn,Freivogel:2005vv}. There are plenty of interesting follow-up directions. The hints for a possible phantom closed Universe within the $Kw$CDM model are not isolated~\citep[e.g.][]{DiValentino:2020hov} and are definitely worth exploring further, particularly in light of the $H_0$ tension and the possible role phantom dark energy alone or in combination with a closed Universe might play in partially reducing this tension, albeit not fully solving it~\citep[see e.g.][for related works]{DiValentino:2016hlg,Zhao:2017cud,Vagnozzi:2019ezj,Visinelli:2019qqu,Alestas:2020mvb,Bose:2020cjb}. It would also be interesting to forecast the role of future cosmic chronometers data in further shedding light on the geometry of the Universe~\citep{Moresco:2015cya,Moresco:2018xdr,Moresco:2020fbm}. And, of course, further research is needed to uncover the origin of the lensing anomaly and the related question of why \textit{Planck} temperature and polarization data appear to prefer a closed Universe, or whether these anomalies are just flukes. These and related questions are left for future work.

\section*{Note added}

The first part of our title takes inspiration from the famous phrase ``Eppur si muove'' (``And yet it moves'' in Italian). This phrase is attributed to Galileo who, after being forced by the Church to retract his claims that the Earth moved around the Sun rather than the other way round, privately defended his claims that the Earth does indeed move. The first part of our title ``Eppur \`{e} piatto''  translates to ``And yet it is flat'' in Italian, obviously referring to the Universe.

\section*{Acknowledgements}

S.V. acknowledges very useful discussions with Eleonora Di Valentino, George Efstathiou, Stefano Gariazzo, Steven Gratton, Will Handley, Raul Jim\'{e}nez, Alessandro Melchiorri, Olga Mena, Seshadri Nadathur, Fabio Pacucci, and Joe Silk, and thanks Cristina Ghirardini for help in producing Fig.~\ref{fig:planck_cc_omegak_age_H0_new}. S.V. is supported by the Isaac Newton Trust and the Kavli Foundation through a Newton-Kavli Fellowship, and acknowledges a College Research Associateship at Homerton College, University of Cambridge. A.L. is partially supported by the Black Hole Initiative at Harvard University, which is funded by grants from the John Templeton Foundation (JTF) and the Gordon and Betty Moore Foundation (GBMF). M.M. acknowledges the grants ASI n.I/023/12/0 and ASI n.2018-23-HH.0. This work was performed using resources provided by the Cambridge Service for Data Driven Discovery (CSD3) operated by the University of Cambridge Research Computing Service (\href{https://www.hpc.cam.ac.uk/}{www.hpc.cam.ac.uk}), provided by Dell EMC and Intel using Tier-2 funding from the Engineering and Physical Sciences Research Council (capital grant EP/P020259/1), and DiRAC funding from the Science and Technology Facilities Council (\href{https://www.dirac.ac.uk/}{www.dirac.ac.uk}).

\section*{Appendix A: impact of cosmic chronometer systematics}
\refstepcounter{appendix}
\label{sec:appendix}

A fundamental step towards reliably estimating $H(z)$ from CC data is to properly take into account systematic effects in the analysis. The various ingredients of the method have to be scrutinized in detail, to establish and quantify possible sources of systematic errors, and take these into account in the total error budget. All these effects have been studied extensively by one of us in~\cite{Moresco:2012by,Moresco:2015cya,Moresco:2016mzx,Moresco:2018xdr,Moresco:2020fbm}, and will be summarized here below.
\begin{enumerate}
    \item Selection of appropriate tracers. Finding an unbiased tracer of the evolution of the differential age of the Universe as a function of redshift is a key step of the analysis. While massive, early, passively-evolving galaxies have been proven to be excellent tracers, in this sense mapping the oldest population of galaxies at each redshift (as we argued in Section~\ref{subsec:cc}), it is important to assess if some residual subdominant young population may bias the result.
    \item Uncertainties in the star formation history (SFH) of the adopted model. Even though the adopted samples consist of extremely old and passive galaxies, these cannot be completely approximated as being simple stellar populations, and it is important to assess in the analysis the impact of considering models with more realistic SFHs.
    \item Uncertainties in the estimated stellar metallicity of the population. This parameter is often used in the analysis as a prior to calibrate the relation to obtain the relative age of a population. It is therefore fundamental to consider in the total error budget also the contribution due to an uncertainty in its estimate.
    \item Dependence on the stellar population synthesis (SPS) model used to calibrate the method. The relative age of a population is obtained through a calibration procedure based on an SPS model. The impact of considering different possible models has to be evaluated accurately to assess its effect on the systematic errors.
\end{enumerate}

\begin{figure}[!t]
\centering
\includegraphics[width=0.7\textwidth]{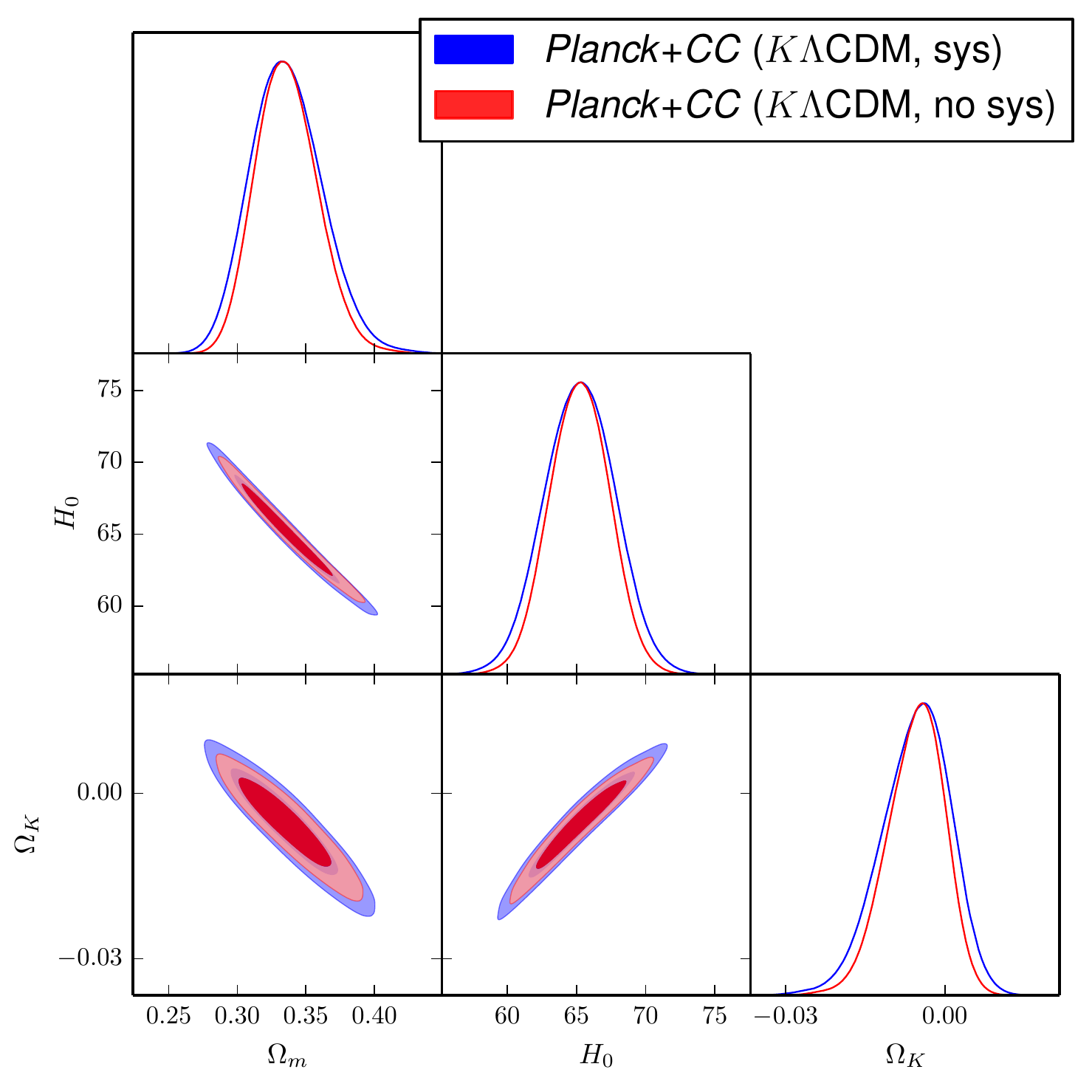}
\caption{Triangular plot showing 2D joint and 1D marginalized posterior probability distributions for $\Omega_m$, $H_0$, and $\Omega_K$ from the \textit{Planck}+\textit{CC} dataset combination, obtained within the 7-parameter $K\Lambda$CDM model, with (blue contours) and without (red contours) the inclusion of SPS systematics: these systematics only broaden the resulting parameter constraints by $\approx 15\%$, leaving all our conclusions qualitatively unchanged.}
\label{fig:compare_systematics}
\end{figure}

We recall here that points 1 to 4 have been carefully evaluated in the past, and that current errors quoted for CC data already take into account in the total error budget the effects from point 1 to 3.

In~\cite{Moresco:2018xdr}, the effect of a possible young subdominant component on $H(z)$ estimates was investigated, and a method to minimize this contamination was proposed, based on a combination of optical and spectroscopic data. A new indicator to quantify the level of contamination based on CaII H and K lines was proposed, and the procedure to propagate this contamination to the total error budget discussed. Current data was found to be compatible with no contamination, as a consequence of the accurate and severe selection cuts applied, following the suggested approach.

In~\cite{Moresco:2012by,Moresco:2015cya,Moresco:2016mzx} the impact of SFH and metallicity uncertainties on the $H(z)$ measurements was extensively discussed, with the associated SFH- and metallicity-related systematics already being included into the total error budget. In particular, the SFH systematics budget was estimated to range between $2\%$ and $3\%$.

The impact of SPS modelling on the results, and further aspects of the impact of metallicity, were studied in detail in~\cite{Moresco:2020fbm}. The mean percentage bias on $H(z)$ as a function of redshift due to different assumptions on SPS models, including adopted stellar library and initial mass function, was quantified in Tab.~3 of~\cite{Moresco:2020fbm}. As this Table shows, the SPS-induced systematic budget is $\lesssim 10\%$, and hence the effect on our results can be expected to be small.

We now assess the impact of these systematics on our results. Recall that, as per our discussion above, the SFH and metallicity contributions to the systematic error budget were already included in the uncertainty of our CC measurements, whereas the sample was found to be compatible with no young subdominant component contamination. Therefore, all that remains to be included is the SPS systematic error budget, which we incorporate following Tab.~3 of~\cite{Moresco:2020fbm}, adding it in quadrature to the statistical uncertainties quoted in Tab.~\ref{tab:cc}. We find that SPS systematics broaden the overall CC uncertainties by typically $\lesssim 10\%$ (and even less for $z \gtrsim 0.9$), which leads to expect that the overall effect on our constraints should be small.

We show the effect of these systematics in the $\Omega_m$-$H_0$-$\Omega_K$ triangular plot in Fig.~\ref{fig:compare_systematics}, where we plot constraints from the \textit{Planck}+\textit{CC} dataset combination within the $K\Lambda$CDM model, both with (blue contours) and without (red contours) SPS systematics included. We see that overall the effect of introducing SPS systematics has been that of slightly broadening the corresponding constraints on $\Omega_m$, $H_0$, and $\Omega_K$ (as well as on all the other cosmological parameters), without shifting the central values. This is expected, given that the SPS systematics have only broadened the CC uncertainties but have not shifted the measurements themselves.

In particular, with [without] systematics we find $\Omega_K = -0.0055 \pm 0.0065$ [$-0.0054 \pm 0.0055$], $H_0 = (65.24 \pm 2.58)\,{\rm km}\,{\rm s}^{-1}\,{\rm Mpc}^{-1}$ [$(65.23 \pm 2.14)\,{\rm km}\,{\rm s}^{-1}\,{\rm Mpc}^{-1}$], and $\Omega_m = 0.336 \pm 0.025$ [$0.336 \pm 0.022$]. In other words, the uncertainties on $\Omega_K$, $H_0$, and $\Omega_m$ have been broadened by approximately $18\%$, $20\%$, and $14\%$ respectively, in line with what we could have expected. As a consequence, the already mild tension with \textit{Planck} within the $K\Lambda$CDM model is further mildened, with $\log{\cal I} \approx -0.43$ compared to the previous $\log{\cal I} \approx -0.47$. Overall these results are therefore extremely stable against observational systematics which affect the CC measurements.

While we have not explicitly tested the impact of these systematics on the results obtained within the extended $Kw$CDM and $M_{\nu}K\Lambda$CDM models, it is completely reasonable to expect that the effect will be similar to that we observed within the $K\Lambda$CDM model, \textit{i.e.} a $\approx 15\%$ broadening of all constraints, which has qualitatively little impact on our conclusions. Our overall conclusion is therefore that the results of our paper are remarkably stable against observational systematics affecting the CC measurements we have adopted.

\bibliography{Curvature_CC}{}
\bibliographystyle{aasjournal}

\end{document}